\documentclass[amsmath,amssymb,nofootinbib,prd,showpacs]{revtex4}

\usepackage{graphicx}
\usepackage{amsmath}
\usepackage{pdfpages}
\usepackage{bm}
\renewcommand{\mathbf}{\bm}
\newcommand{\be}{\begin{equation}}
\newcommand{\ee}{\end{equation}}
\newcommand{\ba}{\begin{eqnarray}}
\newcommand{\ea}{\end{eqnarray}}

\newcommand{\<}{\langle}
\renewcommand{\>}{\rangle}

\renewcommand{\d}{\mathrm{d}}

\newcommand{\so}{{\cal O}}
\newcommand{\sr}{{\cal R}}
\newcommand{\CR}{{\cal R}}

\newcommand{\CD}{{\cal D}}

\renewcommand{\geq}{\geqslant}

\newcommand{\average}[1]{\left\langle #1 \right\rangle_{\CD}}

\begin{document}

\title{The influence of structure formation on the cosmic expansion}

\author{Chris Clarkson, Kishore Ananda and Julien Larena}

\email{chris.clarkson@uct.ac.za, kishore.ananda@gmail.com, julien.larena@gmail.com}

\affiliation{ Cosmology \& Gravity Group, Department of Mathematics
and Applied Mathematics, University of Cape Town, Rondebosch 7701,
Cape Town, South Africa}

\begin{abstract}

We investigate the effect that the average backreaction of structure formation has on the dynamics of the cosmological expansion, within the concordance model. 
Our approach in the Poisson gauge is fully consistent up to second-order in a perturbative expansion about a flat Friedmann background, including a cosmological constant. We discuss the key length scales which are inherent in any averaging procedure of this kind. We identify an intrinsic homogeneity scale that arises from the averaging procedure, beyond which a residual offset remains in the expansion rate and deceleration parameter. In the case of the deceleration parameter, this can lead to a quite large increase in the value, and may therefore have important ramifications for dark energy measurements, even if the underlying nature of dark energy is a cosmological constant. We give the intrinsic variance that affects the value of the effective Hubble rate and deceleration parameter. These considerations serve to add extra intrinsic errors to our determination of the cosmological parameters, and, in particular, may render attempts to measure the Hubble constant to percent precision overly optimistic.

\end{abstract}

\pacs{04.20.-q, 98.80.-k, 98.80.Jk, 91.30.Cd}

\maketitle

\section{Introduction}

The universe appears to be close to homogeneous and isotropic, on average, on large scales, but it exhibits a very clumpy distribution of matter on small scales. To account for this structure, the standard cosmological model relies on the separation of the geometry of spacetime into a perfectly homogeneous and isotropic Friedmann-Lema\^\i tre-Robertson-Walker (FLRW) background describing the large scale properties of the universe, such as the expansion rate, and small fluctuations around this background solution. This provides a straightforward perturbative treatment of the growth of structure under the influence of gravitation. The explicit construction of the background by a smoothing or averaging procedure applied to the clumpy Universe is often ignored, and the background appears as an artificial mathematical object used to perform the calculations of gauge invariant quantities characterising the physical properties of the clumpy universe. 

The essence of this `averaging problem' comes when we try to match the late time universe today, which is full of structure, to the early time universe, which isn't. At the end of inflation we are left with a universe with curvature characterised by some constant $k_\text{inf}$ ($=0,\pm1$ in some units), and cosmological constant, $\Lambda_\text{inf}$, which are fixed for all time (and might be zero), and perturbations which are of tiny amplitude and well outside the Hubble radius; there is no averaging problem at this time, and the idea of background plus perturbations is very natural and simple to define. Fast-forward to today, where structures are non-linear, are inside the Hubble radius, and many have broken away from the cosmic expansion altogether. We may still apparently describe the universe as FLRW plus perturbations to high accuracy; that is, it is natural and seemingly correct to define a FLRW background, \emph{but it is implicitly assumed that this background is the same one that we are left with at the end of inflation}, in terms of $k_\text{inf}$ and $\Lambda_\text{inf}$. Mathematically we can follow a model from inflation to today, but when we try to fit our models to observations to describe our local universe we are implicitly smoothing over structure, and this can contaminate what we think our inflationary background FLRW model should be. Indeed, it is not clear that the background smoothed model should actually obey the field equations at all. Within the standard paradigm, then, the averaging problem also becomes a fitting problem; are the background parameters we are fitting with the CMB actually the same as those when fitting SNIa? (See~\cite{KMM} for a discussion of these issues.)

Because of the non-linearity of the Einstein field equations, the explicit construction of a homogeneous background is far from trivial and it has been know for a long time that the local fluctuations may influence the way the Universe behaves on average~\cite{Ellis84}; this effect is usually dubbed backreaction and has started to be investigated in detail (see, e.g., \cite{backreactionbiblio1,backreactionbiblio2,Rasanen,KMNR,backreactionbiblio3,backreactionbiblio4,backreactionbiblio5,wilt0,wilt1,wilt2,wilt3,backreactionbiblio6,backreactionbiblio7,backreactionbiblio8,lischwarz,lischwarz1,lischwarz2,Iain1,Iain2,Iain3,Kasai,coley1,coley2,AverageArbCoord} and references therein), often as a possible solution to the dark energy problem itself. The problem is mathematical: if we have an inhomogeneous matter distribution in some spacetime, and we try to calculate a homogeneous `background' by smoothing the matter content and calculating the new smoothed metric, we get a different answer than if we smooth the metric directly; this difference is usually termed backreaction, in this context.

In this work, we explore the changes to  the background due to the presence of perturbations by explicitly calculating the effects of backreaction up to second-order in perturbation theory. Similar investigations have been done in the past in the synchronous gauge \cite{lischwarz,lischwarz1,lischwarz2}, and in the Poisson gauge (see e.g. \cite{KMNR,Rasanen,Kasai,Iain1,Iain2}). The authors of these previous works have mainly considered only terms which are quadratic in linear quantities (ignoring the second-order Bardeen potentials), and/or a pure Einstein de-Sitter Universe (ignoring the cosmological constant). 

Here, we present an analysis which is consistent up to second order, both for Einstein de-Sitter and the concordance model~-- flat $\Lambda$CDM~-- as underlying background solutions. To perform our averages we argue that the rest frame of the gravitational field is a natural choice of spatial hypersurface, and that this may be unambiguously defined by the vanishing of the magnetic part of the Weyl tensor, where this is possible. 
We then define the effective expansion rate through the average of the matter fluid expansion projected onto these spatial hypersurfaces, rather than through the expansion 4-scalar for matter (as in \cite{Rasanen,KMNR}) or through the expansion of the `observer' (defined below) flow lines (as in \cite{Kasai,Iain1,Iain2,Iain3}~-- though as we show below, this is in fact the rest frame of the gravitational potential). Our choice can be justified by the fact that the averaging process used in this work is frame (observer) dependent, so that, for a matter of consistency, one wants to retain the observer dependence on the quantities that are averaged also.  On the one hand, the usual expansion 4-scalar for matter is the observed expansion only for observers comoving with the fluid so that it is only appropriate for averaging with respect to a set of comoving observers. On the other hand, the expansion of the observer flow lines is clearly frame dependent, but it does not encompass any information about the matter flow, apart from its background part; the averaged homogeneity being a characteristic of the total matter flow, it seems then more natural to retain a quantity that takes into account the fluctuations in the matter distribution and its peculiar velocity with respect to the observer's rest-frame. 
 
Moreover, we include an important characteristic effect of backreaction, {\it viz.} the existence of an intrinsic variance on quantities such that the Hubble rate and the deceleration parameter (as noted by \cite{lischwarz1,KMNR}). This is important because it gives us our `$1-\sigma$ error' which is an intrinsic unknown when identifying a background based on Gaussian perturbations around FLRW.

The paper is organized as follows. In Section 2 we describe the averaging formalism used in the rest of the paper and based on a recent generalization \cite{AverageArbCoord} of the standard averaging procedure (defined in a comoving coordinate system) to arbitrary coordinate systems. In Section 3, we apply this formalism to the averaging of the cosmological model in the Poisson gauge up to second order. In Section 4, we present the effects on the Hubble rate and deceleration parameter, with a particular emphasis on the importance of the different scales involved (smoothing and averaging) in the process. We find that the effect on the quantities themselves is small (as expected in perturbation theory) but that it can be quite large in terms of the variance affecting these quantities. Our results are quantitatively in agreement with the results obtained in the synchronous gauge \cite{lischwarz,lischwarz1,lischwarz2}. Finally, Section 5 summarizes the results and addresses quickly possible future developements of this work.

\section{Averaging formalism in an arbitrary coordinate system}

In this paper, we are concerned with estimating the backreaction effect in the Poisson gauge of perturbation theory. In this gauge, the 4-velocity of the matter fluid is tilted with respect to the timelike normal vector of the coordinate system, and this makes it difficult to use the standard averaging procedure (see \cite{backreactionbiblio1}), that has been developed with respect to observers comoving with the cosmic matter fluid. A recent work \cite{AverageArbCoord} has generalized the averaging procedure to an arbitrary coordinate system, which allows us to estimate consistently the backreaction effect in the Poisson gauge. This section briefly presents this formalism and discusses the assumptions at the root of it. Essentially, we introduce two velocity fields, one in the rest-frame of the matter content, and another arbitrary one on which we fix an arbitrary family of `observers'. The role of the observers is to introduce spacelike hypersurfaces on which we perform our average; these hypersurfaces are tilted with respect to the spacelike hypersurfaces defined by the fluid (tilted in the sense that their normal vectors are tilted). We fix the coordinates with respect to the observers, and so loosely refer to this as the coordinate frame.

Throughout the paper, we will suppose that gravitation is well described by general relativity on all scales and that the cosmic matter fluid can be considered as a perfect fluid. Moreover, Latin letters of the beginning of the alphabet $(a,b,c,...,h)$ will denote spacetime indices, and Latin letters in the middle the alphabet $(i,j,k,...)$ will denote spatial indices.

\subsection{$1+3$ threading of spacetime}

We consider a set of observers $\so (p)$ defined at each point of the spacetime manifold $p\in {\cal M}$, and characterized by a unit 4-velocity field $n^{a}$ that is everywhere timelike and future directed, i.e. $n^{a}n_{a}=-1$, with zero vorticity. This 4-velocity field induces a natural foliation of spacetime by a continuous set of space-like hypersurfaces locally orthogonal to $n^{a}$. Then we can define the projection tensor field onto these hypersurfaces as $h_{ab}=g_{ab}+n_{a}n_{b}$ that is a well-defined Riemannian metric for these hypersurfaces.
The line element can then be written, with respect to this foliation:
\begin{equation}
\label{eq:ADM_metric}
ds^{2}=-(N^{2}-N_{i}N^{i})\d t^{2}+2N_{i}\d t\d x^{i}+h_{ij}\d x^{i}\d x^{j}\mbox{ ,}
\end{equation}
where we have introduced respectively the lapse function $N(x^{a})$ and the shift 3-vector $N^{i}(x^{a})$ such that the components of the 4-velocity read:
\begin{equation}
\label{eq:rel_velADM}
n^{a}=\frac{1}{N}(1,-N^{i}) \mbox{ , }~~~ n_{a}=N(-1,0,0,0)\mbox{ .}
\end{equation}

For the purposes of this paper, the hypersurfaces orthogonal to $n^{a}$ are characterized by two quantities:
\begin{itemize}
\item the intrinsic curvature of the hypersurfaces: $\sr \equiv h^{ab}\sr_{ab}$, where $\sr_{ab}$ is the 3-Ricci curvature of the hypersurfaces;
\item the extrinsic curvature (or second fundamental form): $K_{ab}\equiv -h^{c}_{a}h^{d}_{b}n_{c;d}$ that encodes the way the hypersurfaces are embedded in the manifold ${\cal M}$.
\end{itemize}

In the following, we will assume that the matter content can be well described by a perfect fluid (not necessarily irrotational) of energy density $\rho(x^{a})$, pressure $p(x^{a})$ and 4-velocity $u^{a}(x^{b})$ (with $u_{a}u^{a}=-1$), so that its stress-energy tensor reads:
\begin{equation}
T_{ab}=(\rho +p)u_{a}u_{b}+pg_{ab}\mbox{ .}
\end{equation}
Note that in this work, the 4-velocity of matter $u^{a}$ is not necessarily aligned with the 4-velocity of the observers $n^{a}$, so that there exists a vector field $v^{a}$ corresponding to the relative velocity of the matter fluid with respect to the fundamental observers. $v^{a}$ is space-like and orthogonal to $n^{a}$ ($v^{a}n_{a}=0$) and one has:
\begin{equation}
\label{eq:relvel}
u^{a}=\gamma (n^{a}+v^{a})~~\mbox{ with }~~ \gamma =\frac{1}{\sqrt{1-v^2}}\mbox{ ,}
\end{equation}
where $\gamma$ is the usual Lorentz factor and $v^2=v^av_a$.
Thanks to the 1+3 foliation, the Einstein field equations can be separated in two different sets: the constraint equations that have to be satisfied on every hypersurface, and the evolution equations, that prescribe how the fields $(h_{ab},K_{ab})$ evolve from one hypersurface to another infinitesimally close.
The  Hamiltonian constraint reads:
\begin{equation}
\label{eq:HC}
\sr -K^{i}_{j}K^{j}_{i}+K^{2}=16\pi G\epsilon+2\Lambda \mbox{ ,~~~where   }~~~~\epsilon=T_{ab}n^{a}n^{b}=\gamma^{2}\rho+(\gamma^{2}-1)p\mbox{ ,}
\end{equation}
where $K\equiv K^{i}_{i}$. The momentum constraint is:
\begin{equation}
\label{eq:MC}
\tilde{\nabla}_{i}K^{i}_{j}-\tilde{\nabla}_{j}K=8\pi G J_{j} \mbox{ ,~~~where~~~ }J_{j}=-T_{ab}n^{a}h^{b}_{j}=\gamma^{2} (\rho+p)v_{j}\mbox{ ,}
\end{equation}
where we have defined the projected covariant 3-derivative on the spatial hypersurfaces of any tensor field $t^{a...c}_{\mbox{ }\mbox{ }\mbox{ }\mbox{ }\mbox{ }d...f}$: $\tilde{\nabla}_{d}t^{a...c}_{\mbox{ }\mbox{ }\mbox{ }\mbox{ }\mbox{ }\bar{a}...\bar{c}}\equiv h^{e}_{d}h^{a}_{a'}...h^{c}_{c'}h_{\bar{a}}^{a''}...h_{\bar{c}}^{c''}\nabla_{e}t^{a'...c'}_{\mbox{ }\mbox{ }\mbox{ }\mbox{ }\mbox{ }a''...c''}$.
The evolution equation for the first fundamental form reads:
\begin{equation}
\label{eq:FFF}
\frac{1}{N}\partial_{t}h_{ij}=-2K_{ij}+\frac{2}{N}\tilde{\nabla}_{(j}N_{i)}\mbox{ ,}
\end{equation}
and for the second fundamental form, one has:
\begin{equation}
\label{eq:SFF}
\frac{1}{N}\partial_{t}K^{i}_{j}=\sr^{i}_{j}+KK^{i}_{j}-\Lambda\delta^{i}_{j}-\frac{1}{N}\tilde{\nabla}_{j}\tilde{\nabla}^{i}N+\frac{1}{N}\left(K^{i}_{k}\tilde{\nabla}_{j}N^{k}-K_{j}^{k}\tilde{\nabla}_{k}N^{i}+N^{k}\tilde{\nabla}_{k}K^{i}_{j}\right)-8\pi G \left(S^{i}_{j}+\frac{1}{2}(\epsilon-S^{k}_{k})\delta^{i}_{j}\right)\mbox{ ,}
\end{equation}
with $S_{ij}=\gamma^{2}\rho v_{i}v_{j}+p(h_{ij}+\gamma^{2}v_{i}v_{j})$.
These equations have to be supplemented by the energy-momentum conservation for the matter fluid: $\nabla_{a}T^{a}_{b}=0$.
We will now introduce the standard decomposition for the covariant spatial derivatives of the 4-vectors in terms of their trace, symmetric trace-free and antisymmetric parts. Writing $\dot{f}\equiv n^{a}\nabla_{a}f$ for any quantity $f$, one has:
\begin{eqnarray}
\nabla_{a}n_{b}&=&-n_{a}\dot{n}_{b}+\frac{1}{3}\xi h_{ab}+\Sigma_{ab}\\
& &\mbox{with }~~~ \xi\equiv h^{c}_{a}h^{da}\nabla_{c} n_{d} \mbox{~~~ and ~~~} \Sigma_{ab}\equiv h^{c}_{a}h^{d}_{b}\nabla_{(c}n_{d)}-\frac{1}{3}\xi h_{ab}\mbox{ ;}\nonumber\\ 
\nabla_{a}u_{b}&=&-n_{a}\left(2\dot{n}_{b}+v\dot{v}n_{b}\right)-\gamma\tilde{\nabla}_{a}v n_{b}+\frac{1}{3}\theta h_{ab}+\sigma_{ab}+\omega_{ab}\\
& & \mbox{with }~~~\theta\equiv h^{c}_{a}h^{da}\nabla_{c} u_{d} \mbox{ ,~~~ } \sigma_{ab}\equiv h^{c}_{a}h^{d}_{b}\nabla_{(c}u_{d)}-\frac{1}{3}\theta h_{ab}\mbox{~~~ and ~~~} \omega_{ab}\equiv h^{c}_{a}h^{d}_{b}\nabla_{[c}u_{d]}\mbox{ ;}\nonumber\\
\nabla_{a}v_{b}&=&-n_{a}\left(\dot{v}_{b}+X_{b}\right)+Y_{a}n_{b}+\frac{1}{3}\kappa h_{ab}+\beta_{ab}+W_{ab}\\
& & \mbox{with }~~~\kappa\equiv h^{c}_{a}h^{da}\nabla_{c} v_{d} \mbox{ ,~~~ } \beta_{ab}\equiv h^{c}_{a}h^{d}_{b}\nabla_{(c}v_{d)}-\frac{1}{3}\kappa h_{ab}\nonumber\\
 & &\mbox{and }~~~ X_{a}\equiv n^{b}h_{a}^{c}\nabla_{c}v_{b}\mbox{ , ~~~} Y_{b}\equiv n^{c}h_{b}^{a}\nabla_{c}v_{a} \mbox{ , ~~~} W_{ab}\equiv h^{c}_{a}h^{d}_{b}\nabla_{[c}v_{d]}\nonumber\mbox{ .}
\end{eqnarray}
In these relations, $\xi$, $\theta$ and $\kappa$ denote the isotropic expansion rates of the 4-velocities $n^{a}$, $u^{a}$ and of the peculiar velocity $v^{a}$ respectively, and $\Sigma_{ab}$, $\sigma_{ab}$ and $\beta_{ab}$ their shears, with respect to the threading of spacetime induced by the vector $n^{a}$. $\omega_{ab}$ and $W_{ab}$ are, respectively, the vorticities\footnote{Since it defines the foliation, $n^{a}$ is vorticity free by definition.} of $u^{a}$ and $v^{a}$ in this same foliation. These quantities are those measured by the observers with 4-velocities $n^{a}$ in their instantaneous rest-frame. In particular $\theta$, $\sigma_{ab}$ and $\omega_{ab}$ differ from the usual expansion, shear and vorticity of the matter fluid as measured by observers comoving with this matter fluid (by acceleration terms essentially), that are defined by the decomposition of $(g^{ac}+u^{a}u^{c})(g^{bd}+u^{b}u^{d})\nabla_{c}u_{d}$. For example, the expansions are linked by the relation:
\begin{equation}
\Theta\equiv \nabla_{a}u^{a}=\theta+\gamma (\gamma^{2}v^{a}\dot{v}_{a}-n^{a}\dot{v}_{a})\mbox{ .}
\end{equation}
Using (\ref{eq:relvel}), one can relate these quantities as follows:
\begin{eqnarray}
\xi&=&\gamma^{-1}\theta-\kappa-\gamma^2 B\\
\Sigma_{ab}&=&\gamma^{-1}\sigma_{ab}-\beta_{ab}-\gamma^2\left(B_{(ab)}-\frac{1}{3}Bh_{ab}\right)\\
W_{ab}&=&\gamma^{-1} \omega_{ab}-\gamma^{2}B_{[ab]}\mbox{ ,}
\end{eqnarray}
where we have introduced the tensor:
\begin{equation} 
B_{ab}\equiv \frac{1}{3}\kappa(v_{a}n_{b}+v_{a}v_{b})+\beta_{ca}v^{c}n_{b}+\beta_{ca}v^{c}v_{b}+W_{ca}v^{c}n_{b}+W_{ca}v^{c}v_{b}\mbox{ ,}
\end{equation}
whose trace is given by $B=\frac{1}{3}\kappa v^{2}+\beta_{ab}v^{a}v^{b}$. In our notation, angular and round brackets denote the antisymmetric and symmetric parts, respectively, of a tensor projected with $h_{a b}$.
Let's finally introduce the following notation for convenience:
\begin{eqnarray}
\theta_{B}&\equiv& -\gamma\kappa-\gamma^{3} B\\
\sigma_{Bij}&\equiv& -\gamma\beta_{ij}-\gamma^{3}\left(B_{(ij)}-\frac{1}{3}Bh_{ij}\right)\mbox{ ,}
\end{eqnarray}
so that:
\begin{eqnarray}
\xi=\gamma^{-1}(\theta+\theta_{B})\mbox{ ,}\\
\Sigma_{ij}=\gamma^{-1}(\sigma_{ij}+\sigma_{Bij})\mbox{ .}
\end{eqnarray}
 
The non-local, free gravitational field is described by the Weyl tensor. Given a timelike vector this is split into electric and magnetic parts. For example, with respect to $n^a$ these are
\be
E_{ab}^{(n)}=C_{acbd}n^cn^d~~~~\text{and}~~~~H_{ab}^{(n)}= {}^*\!C_{acbd}n^cn^d,
\ee
where $C_{abcd}$ is the Weyl tensor and $^*\!C_{abcd}$ is its dual. Analogous definitions exist for the vector field $u^a$. This means that observers in the frame of the fluid and observers in the coordinate frame observe this electric-magnetc split differently (see~\cite{mge} for the transformation relations between the two), analogously to boosted observers measuring different electric and magnetic parts of the electromagnetic field. In particular, in certain gravitational fields there may exist a special frame whereby one of these two components vanishes. For example, in so-called silent universes which are not conformally flat, there exists a preferred frame in which the magnetic part of the Weyl tensor is zero~-- such a frame may be considered the rest-frame of the gravitational field. In spacetimes where this is possible, it is unique as follows from the transformation laws in~\cite{mge}, and there exist (at least) two physical, well motivated, frames: the rest-frame of the fluid, and the rest-frame of the non-local gravitational field. We return to this below.

\subsection{General averaging procedure}

The non-locality of the spatial average that is usually calculated manifests itself problematically when interpreting averaged quantities. What do they mean? In which spacetime do they exist~-- the rough or the smooth?~-- representing physical things (i.e., where are they tensorial objects)? It is tempting to average the fluid expansion of a lumpy spacetime, say, and to think of this as actually being in some sense the `averaged fluid expansion rate'.  Normally, of course, the expansion rate of a fluid is a covariantly defined local object, and so unambiguous when defined in the local rest-frame of the fluid. But it is not understood how to define the rest-frame of the non-local smoothed fluid in a covariant way, and the `average expansion rate' picks up this ambiguity.   

Furthermore, any definition of an averaged expansion rate in the spacetime in which we started is not covariant from a 4-dimensional perspective because the average is with respect to spatial hypersurfaces defined by $n^a$ in the un-smoothed spacetime, and so implicitly rely on some coordinates, and, hence, a mapping between the unsmoothed and smoothed spacetimes.  How do we choose these coordinates? Scalar averaging approaches give us access to some averaged quantities and backreaction terms, \emph{but not the spacetime in which they exist as the objects they are supposed to represent}. Analogously to gauge freedom, this ambiguity leaves implicit choices to make about what objects we consider important, as well as what they mean. Previous analyses have fixed these freedoms in one way or another.

Ideally, we would like to construct a smooth FLRW `spacetime' from an lumpy inhomogeneous spacetime by averaging over structure. This would, in principle, have a metric 
\begin{equation}
\label{EffMetric}
\d s_{\text{eff}}^{2}=-\d\tau^{2}+a_{\CD}^{2}\gamma_{ij}\d y^i\d y^j\mbox{ ,}
\end{equation}
where $\tau$ is the cosmic time and $a_{\CD}(\tau)$ a scale factor, the subscript $\CD$ indicating that it has been obtained at a certain spatial scale characteristic of a compact spatial domain $\CD$, which is large enough so that a homogeneity scale has been reached; in this case $\gamma_{ij}$ will be a metric of constant curvature. This is not to imply that this would be a spacetime in the usual sense because the normal observational relations may not follow directly from this FLRW metric~-- these may have to be calculated separately~\cite{crap,rass}. However, in the context of perturbation theory the effect of renormalisation of the background appears naturally as a first step in this procedure.
 
In the context of averaging a perturbed spacetime we consider below, we can imagine choosing coordinates which straddle both the rough and smooth spacetimes. In particular, we can choose our coordinates in the rough spacetime such that they become the ones we want in the smooth spacetime.
We will choose our coordinates such that the time coordinate in the rough spacetime becomes the proper time coordinate in the smoothed one: that is, we will set $\tau\equiv t$ as well as $y^i=x^i$. (We will also set $N_i=0$.) This effectively de-synchronises the clocks between the rough (with proper time $\int N\d t$ when $N_i=0$) and smoothed (proper time $\tau$) spacetimes. The averaging operator we define is simply the Riemannian average over the domain $\CD$ in the surfaces orthogonal to $n^a$ (i.e., $t=$const.):
\begin{equation}
\label{eq:averager}
\average{\psi}\equiv \frac{1}{V_{\CD}}\int_{\CD}J\d^{3}x\,\psi(t,x^{i})\mbox{ ,}
\end{equation}
and is well defined for any scalar function $\psi$.
This choice has the property that the commutator between partial time derivatives and spatial averages which reads, when $N^i=0$,
\begin{equation}
\label{eq:CommRel}
[\partial_{t}\cdot,\average{\cdot}]\psi(t,x^{i})=\average{N\xi\psi}-\average{N\xi}\average{\psi}\mbox{ ,}
\end{equation}
is zero when we consider perturbed Einstein-de Sitter models below at second-order (that is, the expansion of the spatial hypersurfaces $\xi$ when scaled by $N$ is, for the perturbed model we consider below, $\frac{1}{3}N\xi=H-\dot\Phi-2\Phi\dot\Phi-\frac{1}{2}\dot\Psi^{(2)}$; with $\dot\Phi=0$ as it is for pure dust, the commutator vanishes at second-order).

Of the three expansion rates we have introduced, $\xi$ tells us the expansion of the coordinate grid, and so is not physically attached to the fluid. $\Theta$ measures the fluid expansion rate in its own rest frame, and is the sensible choice of expansion rate with which to characterise the rough spacetime. However, as we have mentioned, after smoothing, the rest-frame of the fluid will change in a way which is not yet known, and will depend on the domain. Any expansion rate we try to investigate must take this into account and so allow for a tilt between the fluid and the normal to the hypersurfaces we use to average. The expansion rate $\theta$, which is the trace of the expansion tensor of the fluid projected into the coordinate rest-frame in which the averaging takes place is the most natural choice when accounting for peculiar velocities in this way. We shall define our observers, which define the spatial hypersurfaces on which the averaging takes place, by the rest frame of the gravitational field; that is, the frame in which $H^{(n)}_{ab}=0$.  At second-order in a perturbation expansion, this is different from the rest-frame of the matter.

Thus, in the rough spacetime, when we consider the length-scale $\ell$ associated with $\theta$, we have 
\[
\frac{1}{3}\theta=n^a\nabla_a\ln\ell=\frac{1}{\ell}\frac{\d \ell}{\d t_\text{prop}}=\frac{1}{N\ell}\frac{\partial \ell}{\partial t}.
\]
Hence, if $t$ represents the proper time in the smoothed spacetime, we can define a pre-synchronised, smoothed, Hubble parameter using $N\theta$ as~\cite{backreactionbiblio2}
\begin{equation}
\label{Eq:Hubble}
H_{\CD}\equiv \frac{1}{3}\average{N\theta}=\frac{1}{3V_{\CD}}\int_{\CD}J\d^{3}x\, N\theta \mbox{ .}
\end{equation}
We may think of this as the average Hubble parameter which preserves the length-scale $\ell$ after smoothing, according to the pre-chosen proper time in the smoothed spacetime. This is referred to as the scaled t-Hubble parameter in~\cite{backreactionbiblio2}, where it is introduced to preserve the structure of the averaged equations. This is not a unique choice and we refer to~\cite{wilt0} for a detailed discussion. In particular, this Hubble parameter changes under a re-scaling of $t$; in the context of perturbations of FLRW, we use $t$ to be the proper time in the background~-- if we were to use conformal time, say, $H_\CD$ would be an averaged conformal Hubble parameter. 
We can use this to then define the effective scale factor for the averaged model as the function $a_{\CD}(t)$ obeying:
\begin{equation}
\label{Eq:SFdef}
H_{\CD}=\frac{\partial_{t}a_{\CD}}{a_{\CD}}\mbox{ .}
\end{equation}
Using the commutation relation, one can average the scalar part of the Einstein field equations to obtain a set of two equations giving the behavior of the effective scale factor $a_{\CD}(t)$ (see \cite{AverageArbCoord} with $N_i=0$):
\begin{eqnarray}
\label{eq:Averaged1}
6H_{\CD}^{2}&=&16\pi G\left(\average{\gamma^{4}N^{2}\rho}+\average{\gamma^{2}(\gamma^{2}-1)N^{2}p}\right)+2\Lambda\average{N^{2}\gamma^{2}}-\average{\gamma^{2}N^{2}\CR}-{\cal Q}_{\CD}+{\cal L}_{\CD}\\
3\frac{\partial^{2}_{t}a_{\CD}}{a_{\CD}}&=&-4\pi G\average{N^{2}\gamma\left(\gamma^{2}\rho(1+v^{2})+3p(1+2\gamma^{2}v^{2})\right)}+\Lambda\average{N^{2}\gamma}\nonumber\\
\label{eq:Averaged2}
 & &+{\cal Q}_{\CD}+{\cal P}_{\CD}+{\cal K}_{\CD}+{\cal F}_{\CD}-{\cal L}_{\CD}\mbox{ ,}
\end{eqnarray}
where we have defined the standard kinematical backreaction:
\begin{equation}
\label{eq:kinback}
{\cal Q}_{\CD}\equiv\frac{2}{3}\left(\average{(N\theta)^{2}}-\average{N\theta}^{2}\right)-2\average{N^{2}\sigma^{2}}\mbox{ ,}
\end{equation}
and additional backreaction terms as:
\begin{eqnarray}
\label{eq:otherback1}
{\cal L}_{\CD}&\equiv&2\average{N^{2}\sigma_{B}^{2}}-\frac{2}{3}\average{(N\theta_{B})^{2}}-\frac{4}{3}\average{N^{2}\theta\theta_{B}}\\
\label{eq:otherback2}
{\cal P}_{\CD}&\equiv&\average{\theta\partial_{t}N}+\average{\gamma N\tilde{\nabla}_{k}\tilde{\nabla}^{k}N}\\
{\cal K}_{\CD}&\equiv&
\average{N^2\gamma^{-1}\theta_{B}
\theta}-3\average{N\gamma^{-1}\theta_{B}
}H_{\CD}\nonumber\\
\label{eq:otherback3}
 & &-\average{N\partial_{t}\theta_{B}}+\average{N^{2}\gamma^{-1}\dot{\gamma}\theta}+\average{N^{2}\gamma^{-1}\dot{\gamma}\theta_{B}} -2\average{N^{2}\theta_{B}^{2}}\\
{\cal F}_{\CD}&=&\frac{2}{3}\average{N^{2}\theta^{2}(\gamma^{-1}-1)}-2\average{N^{2}\sigma^{2}(\gamma^{-1}-1)}-\average{N\theta}\average{N\theta(\gamma^{-1}-1)}\nonumber\\
 & & -\frac{1}{3}\average{N^{2}\theta_{B}^{2}(\gamma^{-1}-1)}-\frac{2}{3}\average{N^{2}\theta\theta_{B}(\gamma^{-1}-1)}-2\average{N^{2}\sigma_{B}^{2}(\gamma^{-1}-1)}\mbox{ .}
\end{eqnarray}

\section{Averaging perturbed FLRW models}

We consider the backreaction effect in a perturbed FLRW model, with a flat background and a cosmological constant. We are interested in the backreaction effect at late times, and so assume the matter in our model is comoving cold dark matter plus baryons. We shall consider scalar modes up to second-order, and ignore vectors and tensors throughout, as the first-order contributions are small and it has been shown that the vectors and tensors induced by first-order scalars are also sub-dominant, although they could provide a slight correction to the results presented here~\cite{MHM,LAC,LACM,ACW,BSTI}. In the Poisson gauge\footnote{In this work, we call a gauge the choice of both the frame, i.e. $n^{a}$, and the coordinate set on hypersurfaces orthogonal to $n^{a}$.} the metric reads~\cite{BMR}
\be
\d s^2=-\left(1+2\Phi+\Phi^{(2)}\right)\d t^2+a^2\left(1-2\Psi-\Psi^{(2)}\right)\delta_{ij}\d x^i\d x^j.
\ee
The first-order scalar perturbations are given by $\Phi, \Psi$, and the second-order by  $\Phi^{(2)}, \Psi^{(2)}$. In this form we have the metric in its Newtonian form, which we may think of as the local rest-frame of the gravitational field.
Comparing with the general metric given above, we are in a gauge with zero shift vector, and $N^2=\left(1+2\Phi+\Phi^{(2)}\right)$ . Moreover, the matter fluid has a peculiar velocity $v^a=(0,v^i)/a$ where  $v_{i}=\frac{1}{2}\partial_{i}(2v^{(1)}+v^{(2)})$, where $v^{(1)}$ is the first order part and $v^{(2)}$ the second order part of the velocity potential. The spatial metric is obviously $h_{ij}=a^{2}\left(1-2\Psi-\Psi^{(2)}\right)\delta_{ij}$. The Poisson gauge is particularly elegant for scalar perturbations because with $n^a$ defined orthogonal to the spatial metric $h_{ij}$, the Weyl tensor becomes
\ba
E^{(n)}_{ij}&=&\frac{1}{2}\left(h_i^{~a}h_{j}^{~b}-\frac{1}{3}h_{ij}h^{ab}\right)\left\{\tilde\nabla_{a}\tilde\nabla_{b}\left[\Phi+\Psi-\Phi^2-\Psi^2+\frac{1}{2}\left(\Phi^{(2)}+\Psi^{(2)}\right)  \right] +\tilde\nabla_a\Phi\tilde\nabla_b\Phi-\tilde\nabla_a\Psi\tilde\nabla_b\Psi  \right\}\\
H^{(n)}_{ij}&=&0.
\ea
In the rest frame $n^a$, then, the gravitational field is silent, and, with $\Psi=\Phi$ is a pure potential field. Hence, $n^a$ may be considered as the rest-frame of the gravitational field, and so defines natural hypersurfaces with which to perform our averages. By contrast, in the frame $u^a$ the Weyl tensor has non-zero $H_{ab}$~\cite{mge}.

As time coordinates we shall use conformal time, $\eta$, proper time in the background, $t$, and background redshift $z=1/a-1$, related by
$a\d\eta=\d t=-\d z/H(1+z)$, interchangeably, where the background Hubble rate $H=\dot a/a=1/a (\d a/\d t)$ is given by
\be
H(z)^2=H_0^2\left[\Omega_0(1+z)^3+1-\Omega_0\right]
\ee
and $\Omega_0=\Omega_m(z=0)$ is the present-day matter density parameter. Note that in the perturbed spacetime the parameter $z$ is just a time coordinate, even though we refer to it as redshift. The Raychaudhuri equation in the background, $\dot H=-\frac{3}{2}H^2\Omega_m$, gives the deceleration parameter 
\be
q_\text{normal}(z)=-\frac{1}{H^2}\frac{\ddot a}{a}=-1+\frac{1+z}{H(z)}\frac{\d H}{\d z}=-1+\frac{3}{2}\Omega_m(z),
\ee
where 
\be
\Omega_m(z)=\frac{\Omega_0(1+z)^3}{\left[\Omega_0(1+z)^3+1-\Omega_0\right]^{1/2}},
\ee
gives the evolution of the density parameter; the density parameter associated with the cosmological constant is $\Omega_\Lambda(z)=1-\Omega_m(z)$.

For a single fluid with zero pressure and no anisotropic stress $\Psi=\Phi$, and $\Phi$ obeys the Bardeen equation
\begin{equation}
\Phi''+ 3 \mathcal{H} \Phi' +  a^2\Lambda  \Phi= 0 =\ddot\Phi+4H\dot\Phi+\Lambda\Phi\,. 
\label{equation of motion for
Bardeen potential}
\end{equation}
and $'=d/d\eta$, and $\mathcal{H}=a'/a$ is the conformal Hubble rate. 
All first-order quantities can be derived from $\Phi$; for example,
\be
v^{(1)}=-\frac{2}{3aH^2\Omega_m}\left(\dot\Phi+H\Phi\right),
\ee
and $\Phi$ is the source of the second-order scalars. The solution to the growing mode of the Bardeen equation may be written as
\be
\Phi(\eta,\bm x)=g(\eta)\Phi_0(\bm x)
\ee
where $\Phi_0(\bm x)$ is the Bardeen potential today ($\eta=\eta_0,~z=0$) and $g(\eta)$ is the growth suppression factor, which may be approximated, in terms of redshift, as~\cite{lahav,Carroll}
\begin{equation}\label{gfac}
g(z) = \frac{5}{2} g_{\infty} \Omega_{m}(z) \left\{
\Omega_{m}(z)^{4/7} - \Omega_\Lambda(z) + \left[ 1 + \frac{1}{2}
\Omega_{m}(z)\right] \left[1 + \frac{1}{70} \Omega_\Lambda(z)
\right] \right\}^{-1}.
\end{equation}
and $g_\infty$ is chosen so that $g(z=0)=1$. 

We define our Fourier transform as (suppressing any temporal quantities)
\be
\Phi(\bm x)=\frac{1}{(2\pi)^{3/2}}\int \d^3k\, \Phi(\bm k)\, e^{i\bm k\cdot\bm x}, 
\ee
where $\Phi^*(\bm k)=\Phi(-\bm k)$. The power spectrum of $\Phi$ is defined  by
\be
\overline{\Phi(\bm k)\Phi(\bm k')}=\frac{2\pi^2}{k^3}\mathcal{P}_\Phi(k)\delta(\bm k+\bm k'), 
\ee
where an overbar denotes an ensemble average. Assuming scale-invariant initial conditions from inflation, this is given by
\be
\mathcal{P}_\Phi(z,k)=\left( \frac{3 \Delta_\mathcal{R}}{5
g_{\infty}} \right )^2 g(z)^2 T(k)^2  
\ee
where $T(k)$ is the normalised transfer function, $\Delta_{\mathcal{R}}^{2}$ is the primordial power of the
curvature perturbation, with~\cite{WMAP5}
$\Delta_{\mathcal{R}}^{2} \approx 2.41 \times 10^{-9}$ at a scale
$k_{CMB}=0.002 \mathrm{Mpc}^{-1}$.

The second-order solutions for $\Psi^{(2)}$ and $\Phi^{(2)}$ are given by~\cite{BMR}. We quote their results directly:
\begin{eqnarray}
\label{PSI}
\Psi^{(2)}(\eta,\bm x)&=&\left( 
B_1(\eta)-2g(\eta)g_{m} -\frac{10}{3}(a_{\rm nl}-1)g(\eta)g_{m}
\right)\Phi_0^2 
\nonumber\\&&
+\left( B_2(\eta) -\frac{4}{3}g(\eta)g_{m}  \right) \Bigg[ \nabla^{-2} 
\left( \partial^i \Phi_0 
\partial_i \Phi_0 \right)- 3 \nabla^{-4} \partial_i \partial^j
\left(\partial^i \Phi_0 \partial_j \Phi_0 \right) \Bigg]\nonumber \\
&&+ B_3(\eta) \nabla^{-2} \partial_i\partial^j(\partial^i \Phi_0 \partial_j 
\Phi_0 )+B_4(\eta) \partial^i \Phi_0 \partial _i\Phi_0 \, ,\\
\label{PHI}
\Phi^{(2)}(\eta,\bm x)&=&\left( B_1(\eta)+4g^2(\eta)
-2g(\eta)g_{m} -\frac{10}{3}(a_{\rm nl}-1)g(\eta)g_{m}
\right)\Phi_0^2 
\nonumber\\&&
+\Bigg[ B_2(\eta)+\frac{4}{3} g^2(\eta) \left( e(\eta)+\frac{3}{2} \right)
-\frac{4}{3}g(\eta)g_{m} \Bigg]  \Bigg[ \nabla^{-2} 
\left( \partial^i \Phi_0 
\partial_i \Phi_0 \right) - 3 \nabla^{-4} \partial_i \partial^j
\left(\partial^i \Phi_0 \partial_j \Phi_0 \right) \Bigg] 
\nonumber\\&&
+B_3(\eta) \nabla^{-2} \partial_i\partial^j(\partial^i \Phi_0 \partial_j 
\Phi_0 )+B_4(\eta) \partial^i \Phi_0 \partial _i\Phi_0\, ,
\end{eqnarray}
where 
$B_i(\eta)={\mathcal H}_0^{-2} \left(f_0+3 \Omega_{0}/2 \right)^{-1} 
\tilde{B}_i(\eta)$ with the following definitions
\begin{eqnarray}
\label{B1B2}
\tilde{B}_1(\eta)&=&\int_{\eta_m}^\eta \d\tilde{\eta} \,{\mathcal H}^2(\tilde{\eta}) 
(f(\tilde{\eta})-1)^2 C(\eta,\tilde{\eta})\, , \,\,\,\,\,\,\,
\tilde{B}_2(\eta)=2\int_{\eta_m}^\eta \d\tilde{\eta} \, {\mathcal H}^2(\tilde{\eta}) 
\Big[2 (f(\tilde{\eta})-1)^2-3
+3 \Omega_m(\tilde{\eta}) \Big] C(\eta,\tilde{\eta})\, , \\
\tilde{B}_3(\eta)&=&\frac{4}{3} \int_{\eta_m}^\eta \d\tilde{\eta} \left(e(\tilde{\eta})
+\frac{3}{2} \right) C(\eta,\tilde{\eta}) \, , \,\,\,\,\,\,\,\,\,\,\,\,
\tilde{B}_4(\eta)= - \int_{\eta_m}^\eta \d\tilde{\eta} \,C(\eta,\tilde{\eta})\, ,
\end{eqnarray}
and 
\begin{equation}
C(\eta,\tilde{\eta})= g^2(\tilde{\eta}) a(\tilde{\eta}) 
\Big[ g(\eta){\mathcal H}(\tilde{\eta})-g(\tilde{\eta}) 
\frac{a^2(\tilde{\eta})}{a^2(\eta)} {\mathcal H}(\eta) \Big] \, ,
\end{equation}
with $e(\eta)=f^2(\eta)/\Omega_m(\eta)$ and
\begin{equation}
f(\eta)=1+\frac{g'(\eta)}{{\mathcal H} g(\eta)}\, .
\end{equation}
$g_m$ denotes the value of $g(\eta_m)$, deep in the matter era before the cosmological constant was important. We also have $a_{\rm nl}$ which denotes any primordial non-Gaussianity present. We shall set this to unity, representing a single field slow-roll inflationary model, in what follows.

We shall also require the perturbed energy density up to second-order
\begin{eqnarray}
\kappa^2   \delta^2 \rho   &=&
\frac{2}{a^2}  \partial^2 {\Psi}^{(2)}   
-6H \dot{\Psi}^{(2)}  -6H^2  {\Phi}^{(2)} 
+24 H^2  \Phi^2  + 6 \dot{\Phi}^2  
+\frac{16}{a^2} \Phi\partial^2\Phi 
\nonumber\\
&&-\frac{8}{3a^2 H^2 \Omega_{m}}
\left[ H^2\left(1-\frac{9}{4}\Omega_m\right)  \partial^{k}{\Phi}\partial_{k}{\Phi}  +2H \partial^{k}{\Phi}\partial_{k}\dot{\Phi}+ \partial^{k}\dot{\Phi}\partial_{k}\dot{\Phi} \right]
\end{eqnarray}
(where $\kappa^2=8\pi G$),
and the Laplacian of the perturbed velocity:
\begin{eqnarray}
3aH^2\Omega_m^2~  \partial^2 \upsilon^{(2)}   
&=&
-{2\Omega_m}\left(  \partial^2\dot{\Psi}^{(2)}   
+H  \partial^2{\Phi}^{(2)}  \right)
\nonumber\\ &&
+4H(\Omega_m-2) \Phi\partial^2\Phi
-4(\Omega_m+2) \dot\Phi\partial^2\Phi
-4(3\Omega_m+2) \Phi\partial^2\dot\Phi
-\frac{8}{H} \dot\Phi\partial^2\dot\Phi
\nonumber\\&&
+4H(\Omega_m-2)\partial^{k}{\Phi}\partial_{k}{\Phi}-16(\Omega_m+1)\partial^{k}{\Phi}\partial_{k}\dot{\Phi}-\frac{8}{H}\partial^{k}\dot{\Phi}\partial_{k}\dot{\Phi}
\nonumber\\&&
+\frac{8}{3a^2 H^2}\left[H \partial^2{\Phi}~\partial^2{\Phi} 
+ \partial^2{\Phi}~\partial^2\dot{\Phi}  +H \partial^{k}{\Phi}~\partial^2\partial_{k}{\Phi}+ \partial^{k}\dot{\Phi}~\partial^2\partial_{k}{\Phi} \right] .
\end{eqnarray}

\subsection{The averaged perturbed EFE}

By making use of the expansion at second order of the three dimensional volume element 
\be
J=a^{3}\left[1-3\Psi+\frac{3}{2}\left(\Psi^{2}-\Psi^{(2)}\right)\right],
\ee
 for any scalar function $\Upsilon$, the Riemannian average $\langle \Upsilon\rangle_{\CD}$ can be expanded in terms of the Euclidean average over the domain $\CD$
 \be
 \langle \Upsilon \rangle =\frac{\displaystyle\int_\CD\d^3x\, \Upsilon }{\displaystyle\int_\CD \d^{3}x}
 \ee
  on the background space slices as:
\begin{equation}
\label{ExpandAverage}
\average{\Upsilon}=\Upsilon^{(0)}+\langle\Upsilon^{(1)}\rangle+\langle\Upsilon^{(2)}\rangle+3\left[\langle\Upsilon^{(1)}\rangle\langle\Psi\rangle-\langle\Upsilon^{(1)}\Psi\rangle\right]\mbox{ ,}
\end{equation}
where $\Upsilon^{(0)}$, $\Upsilon^{(1)}$ and $\Upsilon^{(2)}$ denote respectively the background, first order and second order parts of the scalar function $\Upsilon=\Upsilon^{(0)}+\Upsilon^{(1)}+\Upsilon^{(2)}$. In the rest of the paper, every spatial average will be a Euclidean one.

The averaged Hubble rate as defined by equation (\ref{Eq:Hubble}) is given by: 
\begin{eqnarray}\label{avH}
H_{\mathcal{D}} &=& H -\<\dot{\Phi} \> -\frac{2(1+z)^2}{9  H^2 \Omega_{m}}\left(H\<\partial^2{\Phi} \>+\<\partial^2\dot{\Phi} \>\right)
+\<\Phi~\dot{\Phi}\>\nonumber\\&&
 +\frac{2(1+z)^2}{9 H^3 \Omega_{m}^{2}}\left\{2H\Omega_{m}\left[H\<\Phi~\partial^2\Phi\> +\<\Phi~\partial^2\dot\Phi\>\right]
 +(1+3\Omega_m)H^2\<\partial^{k}{\Phi}~\partial_{k}{\Phi}\>+(2+3\Omega_m)H\<\partial^{k}{\Phi}~\partial_{k}\dot{\Phi}\>
+\<\partial^{k}\dot{\Phi}~\partial_{k}\dot{\Phi}\>
 \right\}
\nonumber\\&&
-3\<{\Phi} \>\<\dot{\Phi} \>
-\frac{2(1+z)^2}{3 H^2 \Omega_{m}}\left[H\<{\Phi} \>\<\partial^2{\Phi} \>+\<{\Phi} \>\<\partial^2\dot{\Phi} \>\right]
\nonumber\\
&&-\frac{1}{2}\< \dot{\Psi}^{(2)} \> +\frac{1}{6}(1+z)\< \partial^2 \upsilon^{(2)} \>.
\end{eqnarray}

The averaged Friedmann equation (\ref{eq:Averaged1}) reads:
\begin{eqnarray}
H_{\mathcal{D}}^2 &=& H^2 
-\frac{1}{6}\left(\mathcal{Q}_{\mathcal{D}}-\mathcal{L}_{\mathcal{D}}+\mathcal{R}_{\mathcal{D}}\right)
-2H\<\dot{\Phi} \> +\frac{2}{3 }(1+z)^2\<\partial^2{\Phi} \>
-4 H^2 \<\Phi^2\> + 2H \<\Phi~\dot{\Phi}\> -\frac{2}{3 }(1+z)^2 \<\Phi~\partial^2\Phi\>
\nonumber\\
&& +\frac{4(1+z)^2}{9 H^2\Omega_{m}^{2} }\left( 1+\Omega_{m}\right)
\left[ H^2 \<\partial^{k}{\Phi}~\partial_{k}{\Phi}\> +2H \<\partial^{k}{\Phi}~\partial_{k}\dot{\Phi}\>+\<\partial^{k}\dot{\Phi}~\partial_{k}\dot{\Phi}\>  \right]
-6H \<{\Phi} \>\<\dot{\Phi} \>
+{2}(1+z)^2\<{\Phi} \>\<\partial^2{\Phi} \>
\nonumber\\
&&+H^2\< {\Phi}^{(2)} \> +\frac{\kappa^2}{6}\< \delta^2 \rho \>,\label{Fried}
\end{eqnarray}

Finally, the averaged acceleration equation (\ref{eq:Averaged2}), which gives an effective Raychaudhuri equation, reads:
\begin{eqnarray}
3\frac{\partial_{t}^{2}a_{\mathcal{D}}}{a_{\mathcal{D}}} &=&
3H^2\left(1-\frac{3}{2}\Omega_{m}\right)
+\mathcal{Q}_{\mathcal{D}}-\mathcal{L}_{\mathcal{D}}+\mathcal{P}_{\mathcal{D}}
+\mathcal{F}_{\mathcal{D}}+\mathcal{K}_{\mathcal{D}}
\nonumber\\&&
+9H^2\left(1-\Omega_{m}\right)\<{\Phi} \>+3H\<\dot{\Phi} \>-(1+z)^2\<\partial^2{\Phi} \>
\nonumber\\
&&
+3H^2\left(9\Omega_{m}-7\right)\<\Phi^2\>-3H\<\Phi~\dot{\Phi}\>+(1+z)^2\<\Phi~\partial^2\Phi\>
\nonumber\\&&
+\frac{(1+z)^2}{3 H^2\Omega_m^2}\left( 4 -9\Omega_{m}\right)
\left[ H^2 \<\partial^{k}{\Phi}~\partial_{k}{\Phi}\> +2H \<\partial^{k}{\Phi}~\partial_{k}\dot{\Phi}\>+\<\partial^{k}\dot{\Phi}~\partial_{k}\dot{\Phi}\>  \right]
\nonumber\\
&&+9H \<{\Phi} \>\<\dot{\Phi} \> +27 H^2 \left(1-\Omega_{m}\right)\<{\Phi} \>^{2}
-{3}(1+z)^2\<{\Phi} \>\<\partial^2{\Phi} \>
\nonumber\\
&&+3H^2\left(1-\frac{3}{2}\Omega_{m}\right) \< {\Phi}^{(2)} \>
-\frac{\kappa^2}{4}\< \delta^2 \rho \>,
\end{eqnarray}

The averaged curvature term is:
\begin{eqnarray}
\mathcal{R}_{\mathcal{D}} &=&2(1+z)^2\left[2\<\partial^2{\Phi} \>+6\<\Phi~\partial^2\Phi\>+3\<\partial^{k}{\Phi}~\partial_{k}{\Phi}\>
+6\<{\Phi} \>\<\partial^2{\Phi} \>+\< \partial^2 {\Psi}^{(2)} \>\right] ,
\end{eqnarray}
and the additional backreaction terms are:
\begin{eqnarray}
\mathcal{F}_{\mathcal{D}} &=&\frac{4(1+z)^2}{3 H^2 \Omega_{m}^{2}}
\left[  H^2 \<\partial^{k}{\Phi}~\partial_{k}{\Phi}\> +2H \<\partial^{k}{\Phi}~\partial_{k}\dot{\Phi}\>+\<\partial^{k}\dot{\Phi}~\partial_{k}\dot{\Phi}\> \right] ,
\\
%
\mathcal{P}_{\mathcal{D}} &=& 3H\<\dot{\Phi} \>+(1+z)^2\<\partial^2{\Phi} \>-15H\<\Phi~\dot{\Phi}\>-3\<\dot{\Phi}^2 \>
-(1+z)^2\left[\<\Phi~\partial^2\Phi\>+2\<\partial^{k}{\Phi}~\partial_{k}{\Phi}\>\right]
\nonumber\\
&&-\frac{2(1+z)^2}{3 H^2 \Omega_{m}}\left[H \<\dot\Phi~\partial^2\Phi\>+\<\dot\Phi~\partial^2\dot\Phi\>  \right]
+9H\<{\Phi} \>\<\dot{\Phi} \>+3(1+z)^2\<{\Phi} \>\<\partial^2{\Phi} \>
\nonumber\\
&&
+\frac{1}{2}(1+z)^2\< \partial^2 {\Phi}^{(2)} \>+\frac{3}{2}H\< \dot{\Phi}^{(2)} \> ,
%
%
\\
\mathcal{Q}_{\mathcal{D}}-\mathcal{L}_{\mathcal{D}} &=&
\frac{8(1+z)^2}{3  H \Omega_{m}}
\left[H \<\partial^2{\Phi} \> +\<\partial^2\dot{\Phi} \>\right] + 6\<\dot{\Phi}^2 \> - 6\<\dot{\Phi} \>^{2}
\nonumber\\
&&-\frac{8(1+z)^2}{3 H \Omega_{m}}
\left[2H\<\Phi~\partial^2\Phi\>+2\<\Phi~\partial^2\dot\Phi\> +3H\<\partial^{k}{\Phi}~\partial_{k}{\Phi}\>+3\<\partial^{k}{\Phi}~\partial_{k}\dot{\Phi}\>
\right] 
\nonumber\\
&& -\frac{8(1+z)^4}{27 H^4 \Omega_{m}^{2}}
\left[H^2\<\partial^2{\Phi} \>^{2}+2H\<\partial^2{\Phi} \>\<\partial^2\dot{\Phi} \>+\<\partial^2\dot{\Phi} \>^{2} \right]
\nonumber\\
&&-\frac{8(1+z)^2}{3 H^2 \Omega_{m}^{2}}
\left[-3H^2\<{\Phi} \>\<\partial^2{\Phi} \> -3H\<{\Phi} \>\<\partial^2\dot{\Phi} \> +H\<\dot{\Phi} \>\<\partial^2{\Phi} \>
+\<\dot{\Phi} \>\<\partial^2\dot{\Phi} \>\right]
\nonumber\\
&&-2H(1+z)\< \partial^{2}{\upsilon}^{(2)} \> ,
%
%
%
\\
\mathcal{K}_{\mathcal{D}} &=&\frac{(1+z)^2}{ H^2 \Omega_{m}}\Bigg\{
\frac{4H}{3 }\left[H\left(1-\frac{3}{4}\Omega_m\right)\<\partial^2{\Phi} \>+\<\partial^2\dot{\Phi} \>  \right]
\nonumber\\
&&-\frac{2}{3 }\left[H^2\left({4}-3\Omega_m\right)\<\Phi~\partial^2\Phi\>+4H\<\Phi~\partial^2\dot\Phi\>+3H\<\dot\Phi~\partial^2\Phi\>+3\<\dot\Phi~\partial^2\dot\Phi\>\right]
\nonumber\\
&&+\frac{1}{3 \Omega_{m}}\left[
3H^2(3\Omega_m^2-2\Omega_m-4)\<\partial^{k}{\Phi}~\partial_{k}{\Phi}\>
-8H\<\partial^{k}{\Phi}~\partial_{k}\dot{\Phi}\>
-2(2-3\Omega_m)\<\partial^{k}\dot{\Phi}~\partial_{k}\dot{\Phi}\>
\right]
\nonumber\\
&&-\frac{4(1+z)^2}{3 H^2 \Omega_{m}}
\left[H^2 \<\partial^2{\Phi}~\partial^2{\Phi}\> +2H\<\partial^2{\Phi}~\partial^2\dot{\Phi}\>+\<\partial^2\dot{\Phi}~\partial^2\dot{\Phi}\>\right]
\nonumber\\
&&+\left[ 
H^2\left(4-3\Omega_{m}\right)\<{\Phi} \>\<\partial^2{\Phi} \>+2\<\dot{\Phi} \>\<\partial^2\dot{\Phi} \>+2H\<\dot{\Phi} \>\<\partial^2{\Phi} \>
+4H\<{\Phi} \>\<\partial^2\dot{\Phi} \>
\right]
\nonumber\\
&&+\frac{4(1+z)^2}{9 H^2 \Omega_{m}}\left[H^2\<\partial^2{\Phi} \>^{2}
+2H\<\partial^2{\Phi} \>\<\partial^2\dot{\Phi} \>+\<\partial^2\dot{\Phi} \>^{2}\right]\Bigg\}
\nonumber\\
&&
+\frac{1}{2}(1+z)\< \partial^{2}\dot{\upsilon}^{(2)} \> -\frac{1}{2}H(1+z)\< \partial^{2}{\upsilon}^{(2)} \> .
\end{eqnarray}

\subsection{The averaging procedure}

We have expanded all perturbative quantities in terms of a Euclidean average, which will let us calculate the averaged quantities of interest in terms of the primordial power spectrum of $\Phi$.
Our Euclidean averaging procedure follows~\cite{KMNR} which we summarise here. We apply our spatial average over a spherical domain using a finite window function, and then we apply an ensemble average to tell us what this spatial average will be in a typical domain. We may also calculate the variance we may expect as we move from domain to domain, which we would expect to vanish as the domain size becomes large. 

The spatial average involves a specific domain $\cal D$ which is arbitrary in principle. Here we use a window function $W$ to specify the domain size and shape. For any spatial variable $X(\bm x)$, let
\be
\< X(\bm x)\>=\frac{1}{V}\int \d^3 x\, W(x/R_\mathcal{D})\, X(\bm x)
\ee
where the integral is taken over all space, and $R_\mathcal{D}$ specifies the size of the domain. We shall take our domain to be spherical as we expect a roughly isotropic distribution, and use a Gaussian window function for simplicity: $W(y)=e^{-y^2/2}$, so that 
\be
V=\int \d^3 x\, W(x/R_\CD)=4\pi R_\CD^3\int_0^\infty y^2W(y) \d y = (2\pi)^{3/2} R_\CD^3
\ee
for any $R$. Here we use 
\be
W(kR_\CD)=\frac{1}{V}\int\d^3x W(x/R_\CD) \, e^{-i\bm k\cdot\bm x}
\ee
as the Fourier transform of $W(x/R_\CD)/V$. A top-hat window function could also be used instead.

Without a specific realisation of perturbations to work from we must calculate what the average of $X(\bm x)$ would be in a typical domain of size $R_\mathcal{D}$. Primordial perturbations are usually taken to be Gaussian fluctuations with zero mean, which implies that we can calculate what the ``average of $X(\bm x)$ would be in a typical domain of size $R_\mathcal{D}$" by taking an ensemble average over many domains~\cite{LL,Ruth}. Following \cite{KMNR} we denote this separate, additional average by an over-bar.

It is clear that once the ensemble average is taken in our averaged backreaction equations, all stand-alone first-order perturbations will vanish if we assume that $\Phi(\bm x)$ is a Gaussian variable, and only quadratic -- second-order -- quantities will remain.  Let us therefore apply this spatial-then-ensemble average to the product of two Gaussian random scalars $A(\bm x)$ and $B(\bm x)$. We find first that
\be
\<A(\bm x)\, B(\bm x)\>=\frac{1}{(2\pi)^3}\int \d^3 k_1\int \d^3 k_2\,\, W(|\bm k_1+\bm k_2|R_\mathcal{D}) A(\bm k_1)\, B(\bm k_2).
\ee
Now, if $A$ and $B$ are statistically homogeneous, then we may write $A(\bm k)=A(k)E(\bm k)$ where $E$ is a unit random variable satisfying $E^*(\bm k)=E(-\bm k)$ and
\be
\overline{E(\bm k_1) E(\bm k_2)}=\delta(\bm k_1+\bm k_2).
\ee
Assuming that $A$ and $B$ are perfectly correlated random variables, we then find
\be
\overline{\<A(\bm x)\, B(\bm x)\>}= \frac{1}{(2\pi)^3}\int\d^3 k\, A(k) B(k)\, .
\ee 
Note that this is a constant as we'd expect, and that the window function has dropped out of the average. That is, once a statistical average is taken, the size of the domain doesn't make any difference for the product of Gaussian Random fields.  We return to this below. Similarly we find
\be
\overline{\<\partial_iA(\bm x)\, \partial^iB(\bm x)\>}= + \frac{1}{(2\pi)^3}\int\d^3 k\, k^2 A(k) B(k)\, .
\ee

We shall also require ensemble averages of the product of spatially averaged  variables, which arise from commuting the spatial average with the time derivative. For example
\be
\overline{\<A(\bm x)\>\< B(\bm x)\>}= \frac{1}{(2\pi)^3}\int\d^3 k\, W(kR_\mathcal{D})^2\,A(k) B(k)\, ,
\ee
which does depend on the domain size~-- it is just $\overline{\<A(\bm x)\, B(\bm x)\>}$ with a squared window function stuck in the integral.

Most of the terms we are dealing with are scalars schematically of the form $\partial^m\Phi(\bm x)\partial^n\Phi(\bm x)$ where $m$ and $n$ represent the number of derivatives (not indices), such that $m+n$ is even so that there are no free indices. (For example, $\partial_i\Phi\partial^2\partial^i\Phi$ has $m=1$ and $n=3$.) Then
\be\label{av-phi}
\overline{\<\partial^m\Phi(\bm x)\partial^n\Phi(\bm x)\>}
=(-1)^{(m+3n)/2}\int \d k\, k^{m+n-1}\mathcal{P}_\Phi(k).
\ee
In particular, we then also find
\be\label{av-lapphi}
\overline{\<\partial^2[\partial^m\Phi(\bm x)\partial^n\Phi(\bm x)]\>}
=0.
\ee
Time derivatives of the Bardeen potential may be dealt with using $\displaystyle\dot\Phi(t,\bm x)=-\frac{1}{H}\frac{\d \ln g}{\d z}\Phi(t,\bm x)$.

\subsubsection{Ensemble Average of Inverse Laplacian Terms}

In the solutions for the second-order Bardeen potentials, we encounter terms involving inverse Laplacians of quadratic first-order variables, such as
\be
\partial^{-2}(\partial_i\Phi\partial^i\Phi)~~~\mbox{and}~~~\partial^{-2}\partial_i\partial^j[\partial^i\Phi\partial_j\Phi].
\ee
These need some care when we take the ensemble average.

We need to find the ensemble average of an object which is the inverse Laplacian of a product of Gaussian Random Fields. Consider
\be\label{invlap}
\partial^2 X(\bm x)=A(\bm x)B(\bm x)
\ee
so that we may write, formally, 
\be
X(\bm x)=\partial^{-2} [A(\bm x)B(\bm x)]. 
\ee
(Note that $X(\bm x)$ itself is not a product of Gaussian random fields since $\overline{\partial^2 [A(\bm x)B(\bm x)]}=0$~-- i.e., the Laplacian of the product of Gaussian random fields has zero mean~-- while $\overline{\partial^2 X(\bm x)}\neq0$.)  In Fourier space we have
\be
X(\bm k)=-\frac{1}{(2\pi)^{3/2}}k^{-2}\int\d^3k' A(\bm k') B(\bm k-\bm k').
\ee
Taking the ensemble average of this equation gives us a $\delta(\bm k)/k^2$ term; if we are to find $\overline{X(\bm x)}$ then we would have to integrate over $k$-space which now involves this awkward divergent term. Let us start instead from the formal solution to Eq.~(\ref{invlap}) in real space which is, ignoring boundary terms and any homogeneous solution,
\ba
X(\bm x)&=&-\frac{1}{4\pi}\int \frac{d^3 x'}{{|\bm x -\bm x'|}}{A(\bm x')B(\bm x')}\\
&=&-\frac{1}{4\pi}\frac{1}{(2\pi)^3}\int \frac{d^3 x'}{|\bm x -\bm x'|}\int d^3k_1\int d^3k_2\,  A(\bm k_1)B(\bm k_2)e^{i(\bm k_1+\bm k_2)\cdot \bm x'}.
\ea
Now, if $A$ and $B$ are Gaussian Random Fields, then we have 
\be\label{ensemble-av-invlap}
\overline{X(\bm x)}= -\frac{1}{4\pi}\frac{1}{(2\pi)^3}\int \frac{d^3 x'}{|\bm x -\bm x'|}\int d^3k\, A(k)B(k).
\ee 
That is, the inverse Laplacian gives a divergent $\int \d^3 x/x$ factor.
To relate to Fourier space, we can use the relation
\be
\frac{1}{|\bm x -\bm x'|}=\frac{1}{2\pi^2}\int d^3 k\frac{e^{i\bm k\cdot(\bm x-\bm x')}}{k^2},
\ee
so that we have 
\be
\overline{X(\bm x)}=-\frac{1}{(2\pi)^3}\int d^3k\int d^3k'\, \frac{ \delta(\bm k)}{ k^2} A(k')B(k') e^{i\bm k\cdot\bm x}, 
\ee
which is what we get if we start in Fourier space.\footnote{Using $\displaystyle\delta(\bm k)=\frac{1}{(2\pi)^3}\int\d^3 y e^{i\bm k\cdot\bm y}$ we find the identity $\displaystyle\int d^3k \frac{\delta(\bm k)}{k^2} e^{i\bm k\cdot\bm x}=\frac{1}{4\pi}\int\frac{\d^3y}{|\bm x+\bm y|}$.} (Despite appearances this isn't actually a function of $x$ -- we can expand out the $k$ integral into angular and radial parts with the $k_z$-axis parallel to $\bm x$ to see this.) It is clear that calculating the ensemble average of an inverse Laplacian using Eq.~(\ref{ensemble-av-invlap}) will be easier.

Now we must find the ensemble averages of $\overline{\partial^{-2}\partial_i\partial^j[\partial^i\Phi\partial_j\Phi]}$, and $\overline{\partial^{-4}\partial_i\partial^j[\partial^i\Phi\partial_j\Phi]}$. Let us investigate
\be
\partial^{-n}\left\{\partial_i\partial_j[\partial^iA(\bm x)\partial^j B(\bm x)]\right\}=\frac{(-1)^{n/2}}{(2\pi)^3}\int\d^3k_1\int\d^3k_2 A(\bm k_1)B(\bm k_2-\bm k_1)f(\bm k_1,\bm k_2)e^{i\bm k_2\cdot \bm x}
\ee
where $n=2$ or 4 and
\be
f(\bm k_1,\bm k_2)=\frac{(\bm k_1\cdot\bm k_2)(\bm k_2-\bm k_1)\cdot\bm k_2}{k_2^n}.
\ee
If we try to take the ensemble average at this stage we get $\delta(\bm k_2)/k_2^n$ as well as mixed arguments in $A$ and $B$ (which means we can't easily perform the angular part of either integral). We may however write $B(|\bm k_2-\bm k_1|)\delta(\bm k_2)=B(k_1)\delta(\bm k_2)$ and perform the angular part of the $\bm k_1$-integral, which gives \footnote{This is a bit ambiguous, however, because $f(\bm k_1,\bm k_2)$ is singular at $k_2=0$ (although doing this does give the correct answer, provided $B$ is well behaved). For example, we could take some of $f$ into this and write $B(|\bm k_2-\bm k_1|)(\bm k_1\cdot\bm k_2)\delta(\bm k_2)=B(k_1)(\bm k_1\cdot\bm 0)\delta(\bm k_2)=0$, which gives the wrong answer. To be more precise, let us write 
\be\label{trick}
A(\bm k)=\int\d^3k'A(\bm k-\bm k')\delta(\bm k')
\ee 
so that we have 
\be
\partial^{-n}\partial_i\partial_j[\partial^iA(\bm x)\partial^j B(\bm x)]=\frac{(-1)^{n/2}}{(2\pi)^3}\int\d^3k_1\int\d^3k_2\int\d^3k_3 A(\bm k_1-\bm k_3)B(\bm k_2-\bm k_1)f(\bm k_1,\bm k_2)\delta(\bm k_3)e^{i\bm k_2\cdot \bm x}
\ee
and now, taking the ensemble average,
\ba
\overline{\partial^{-n}\partial_i\partial_j[\partial^iA(\bm x)\partial^j B(\bm x)]}&=&\frac{(-1)^{n/2}}{(2\pi)^3}\int\d^3k_1\int\d^3k_2\int\d^3k_3 
A(|\bm k_1-\bm k_3|)B(|\bm k_2-\bm k_1|)f(\bm k_1,\bm k_2)\delta(\bm k_3)\delta(\bm k_2-\bm k_3)e^{i\bm k_2\cdot \bm x}\nonumber\\
&=&
\frac{(-1)^{n/2}}{(2\pi)^3}\int\d^3k_3\int\d^3k_4 A(k_4)B(k_4)f(\bm k_3-\bm k_4,\bm k_3)\delta(\bm k_3) e^{i\bm k_3\cdot \bm x}
\ea
where the $k_2$ integral was done and $\bm k_4=\bm k_3-\bm k_1$. We now perform the angular part of the $k_4$ integral, giving the result stated in the text.}
\ba
\overline{\partial^{-n}\partial_i\partial_j[\partial^iA(\bm x)\partial^j B(\bm x)]}&=&-\frac{2}{3}\frac{(-1)^{n/2}}{(2\pi)^2}\int\d^3k' \frac{\delta(\bm k')}{k'^{n-2}}e^{i\bm k'\cdot\bm x}\int_0^\infty\d k\,\,k^4A(k)B(k) \\
&=& \frac{1}{6\pi^2}\int_0^\infty \d k\,\, k^4 A(k)B(k)~~~~\mbox{if}~~~~n=2\\
&=& -\frac{1}{3(2\pi)^3}\int\frac{\d^3\bm x'}{|\bm x-\bm x'|}\int_0^\infty \d k\,\, k^4 A(k)B(k)~~~~\mbox{if}~~~~n=4.
\ea
Hence, for $A=B=\Phi$, we find
\ba
\overline{\<\partial^{-2}\partial_i\partial_j[\partial^i\Phi(\bm x)\partial^j \Phi(\bm x)]\>}&=& \frac{1}{3}\int_0^\infty \d k\,  k \mathcal{P}_\Phi(k)
=\frac{1}{3}\overline{\<\partial_i\Phi(\bm x)\partial^i\Phi(\bm x)\>},
\\
\overline{\<\partial^{-4}\partial_i\partial_j[\partial^i\Phi(\bm x)\partial^j \Phi(\bm x)]\>}&=& -\frac{1}{12\pi}\left\<\int\frac{\d^3\bm x'}{|\bm x-\bm x'|}\right\>\int_0^\infty \d k\, k  \mathcal{P}_\Phi(k)
\nonumber\\
&=&-\frac{1}{12\pi}\left\<\int\frac{\d^3\bm x'}{|\bm x-\bm x'|}\right>\overline{\<\partial_i\Phi(\bm x)\partial^i\Phi(\bm x)\>}.
\ea
Formally, for a scale-invariant spectrum the latter of these is infinite. However, where this appears in the second-order Bardeen potentials, it actually doesn't contribute, because we have shown the important result:
\be
\overline{ \partial^{-2} 
\left( \partial^i \Phi 
\partial_i \Phi \right)- 3 \partial^{-4} \partial_i \partial^j
\left(\partial^i \Phi \partial_j \Phi \right)}=0.
\ee
That is, if $\partial^4X(\bm x)=\partial^i\partial^j\left[\partial_{\<i}\Phi(\bm x)\partial_{j\>}\Phi(\bm x)\right]$ then $\overline{X(\bm x)}=0$ (angle brackets on the indices denote symmetric trace free part of a tensor), which also says that the ensemble average of the scalar part of $\partial_{\<i}\Phi\partial_{j\>}\Phi$ is zero.
Hence, after ensemble averaging, the second-order Bardeen potentials have no divergent terms in the IR from inverse Laplacians.
Similarly, using these results it is possible to show that 
\be
\overline{\<\partial^2\Psi^{(2)}(\bm x)\>}=\overline{\<\partial^2\Phi^{(2)}(\bm x)\>}=\overline{\<\partial^2v^{(2)}(\bm x)\>}=0.
\ee

\subsection{Length scales, smoothing and divergences}

There are two separate divergences which occur in the averaged equations. We have a UV divergence from terms like $\int\d k k^{\alpha}\mathcal{P}_\Phi$, and IR divergences from the same types of terms. We discuss these in turn.

\subsubsection{UV divergences}

On small scales the transfer function behaves as $(\ln k)/k^2$ for pure CDM, and so integrals such as Eq.~(\ref{av-phi}) diverge whenever $m+n\geq 4$. Recall that this is independent of the size of the domain which has already been taken into account. There are different ways to deal with this divergence. If inflation introduces a tilt to the primordial power spectrum then the divergence only formally occurs for $m+n> 4$~-- that is, in terms which don't appear here. However, in the case $m+n=4$ for a tilt of $n_s=0.95$, for example,  convergence of the integrals only happens for $k/k_{eq}>10^{20}$ which is absurdly tiny scales -- so this is effectively a divergence. Secondly, if the baryon fraction is high enough, this cuts off small-scale power and can remove some of the divergences. But we should expect finite results for basic CDM too (though of course in reality there is a cutoff at tiny galactic scales). Alternatively,  we may introduce a small-scale cutoff by hand in the Fourier integrals. The corresponding sampling in real space is then oscillates a bit like $x\sin(1/x)$ -- which is clearly not suited for the purpose of averaging. 

Instead, we shall remove the UV divergences which appear by smoothing our perturbation $\Phi(\bm x)$ in real space before averaging it. Since linear theory is only accurate up to some scale and under-estimates power below this scale, it makes sense to smooth away structure on scales smaller than this before we apply our averaging procedure~-- presumably higher-order perturbation theory will be required to correct for this. We replace $\Phi$ with a weighted average over nearby points using~\cite{LL}
\be
\Phi_{\mathcal{S}}(\bm x)=\frac{1}{V_{\mathcal{S}}}\int\d^3 x'\, W(|\bm x'-\bm x|R_\mathcal{S}) \,\Phi(\bm x'),
\ee  
where for simplicity we use the same window function as we use in calculating the averages, but we separate the smoothing length scale $R_\mathcal{S}$ from the averaging length scale $R_\mathcal{D}$. In Fourier space this amounts to replacing $\Phi(k)$ everywhere with $W(kR_\mathcal{S})\Phi(k)$, and $\mathcal{P}_\Phi$ with $W(kR_\mathcal{S})^2\mathcal{P}_\Phi$. Note that the only terms which are made finite by this smoothing are the terms with four derivatives in them such as $\overline{\<\partial^2\Phi\partial^2\Phi\>}$ -- although all terms are affected to some degree. These terms appear in the generalised Raychaudhuri equation, but not in the averaged Hubble rate. Hence, this divergence can significantly affect the effective equation of state, but not the averaged expansion rate. On the other hand, there are terms in both the Friedmann and Raychaudhuri equations like $\overline{\<\partial^2\Phi\>\<\partial^2\Phi\>}$ which have the same kind of divergence as $R_\mathcal{D}\to0$, but these divergences are controled by the domain size.

\subsubsection{IR divergences}

While the UV divergences are hopefully the result of using second-order perturbation theory and so can be removed by hand until a fully relativistic renormalised approach is available, the IR divergences are maybe a much deeper issue (see, e.g.,~\cite{lyth} for a discussion). For pure dust the integrals converge as $k\to0$ (the only divergent terms are $\<\Phi^2\>$ which cancel in the Friedmann equation); this IR divergence shows up in the variance of the expansion rate~\cite{KMNR}. For $\Lambda$CDM, however, this doesn't happen because $\dot\Phi\neq0$ and many terms appear which have an IR divergence. For a scale-invariant primordial power spectrum the IR divergence is only logarithmic, so is not really an issue. They may be removed by inserting a Hubble scale cutoff in the lower limits of the $k$-integrals~-- but why choose the Hubble scale? 

During matter domination the Hubble or horizon scale is not really of direct physical significance, other than a symbolic `size' of the universe. As far as structure formation goes it plays no role at all at this perturbative level. The only physical scale (crudely speaking) is the equality scale $k_{eq}$ which is the Hubble scale at matter-radiation equality. The longest wavelength modes in the present-day universe are set by the first modes leaving the Hubble radius at the beginning of inflation, and not by the modes which happen to be entering the horizon today. Modes longer than the present day horizon are effectively homogeneous and so are usually considered to be part of the background, effectively renomalising them away. However, this argument only works if those super-Hubble wavelength modes, when renormalised to be part of the background, enter the field equations as curvature terms. \emph{This is not the case for backreaction.} The time evolution of the averaged quantities is complicated and doesn't scale simply as $1/a^2$. Hence, such modes may be considered part of the homogeneous background, but not in a theory of gravity obeying the usual Einstein field equations. Indeed, it is exactly that homogeneous background that the backreaction approach aims to calculate. 

It is not clear whether modes longer than the Hubble radius are real~-- if the standard inflationary picture is correct~-- and might have to be taken into account in the averaging backreaction approach. If the super-Hubble spectrum is tilted to the red this could be important~\cite{KMNR}; conceivably back-reaction may allow a mechanism to probe this part of the power spectrum.  We set the lower IR limit of $k$-integration to be $k_{H}/L$ where $L$ is the largest wavelength mode we are prepared to include, in units of the present day Hubble scale. For simplicity we just assume a scale-invariant spectrum for the whole range of $L$, and fix $L$ to be ten times larger than the Hubble scale. Generally we get the same answers as if we fix it at the Hubble scale, but this allows us to have a domain as large as the Hubble radius.

\section{The averaged Hubble rate and deceleration parameter}

We shall use length scales intrinsic to the model as reference points for smoothing and averaging: the Silk scale, $k_\text{silk}^{-1}$ the equality scale, $k_\text{eq}^{-1},$ and the Hubble scale, $k_\text{H}^{-1}$, and so the baryon fraction appears as this governs the Silk scale. Recall that~\cite{EH}
\ba
k_\text{silk}&\approx& 1.6\, \left( \Omega_{{b}}{h}^{2} \right) ^{ 0.52} \left( \Omega_{{0}}
{h}^{2} \right) ^{ 0.73} \left[ 1+ \left(10.4 \Omega_{{
0}}{h}^{2} \right) ^{- 0.95} \right] 
\text{Mpc}^{-1}, \\
k_\text{eq}&\approx& 7.46 \times 10^{-2}\Omega_0h^2 \text{Mpc}^{-1}, ~~~\text{and}~~~ k_\text{H}=\frac{h}{3000}\text{Mpc}^{-1},
\ea 
where $\Omega_b$ and $\Omega_0$ are the baryon and total matter contributions today and $H_0=100 \,h\,$kms$^{-1}$Mpc$^{-1}$.

We shall use two models for comparison: Einstein-de Sitter with $h=0.7$ and 5\% baryon fraction (WMAP5~\cite{WMAP5} estimates $\Omega_b\approx0.046$). This has $k_ \text{eq}^{-1}\simeq 27.9$Mpc and $k_ \text{silk}^{-1}\simeq 6.0$Mpc. The other model we shall use is a concordance model with $\Omega_0=0.26, h=0.7, f_\text{baryon}=0.175$ (this is the WMAP5 best fit~\cite{WMAP5}). The key length scales in this model are $k_ \text{eq}^{-1}\simeq 107.2$Mpc and $k_ \text{silk}^{-1}\simeq 11.5$Mpc. Both models have $k_ \text{H}^{-1}\simeq4.3$Gpc. To calculate the integrals we use transfer functions presented in~\cite{EH}. All lengths shown are in Mpc.

We set $L=10$; that is, all $k$-integrals have an IR cut-off set at ten times the Hubble scale.

\subsection{Hubble rate}

There are different aspects of the backreaction we wish to probe, and some subtleties arise because we have to take an ensemble average of our equations. When we examine the Hubble rate we are interested in two things: the dynamics of the expansion rate, and the averaged Friedmann equation. In the Friedmann equation we are interested in quantifying the new terms which enter the field equations as a result of averaging, which are the new components which drive the expansion; within the context of dark energy, it is common to think of these as effective fluid or curvature terms. While the spatial average of these two agree up to perturbative order, when we take the ensemble average we ascertain different information. 

Take the ensemble average of the Hubble rate given by the generalised Friedmann equation. For a given domain size this tells us the expectation value of the averaged Hubble rate we might expect to find dynamically in the field equations. When we present $H_\mathcal{D}$ below, we have usually calculated 
\be
\breve{H}_\mathcal{D}\equiv\sqrt{\overline{H_\mathcal{D}^2}}
\ee
 i.e., we have taken the ensemble average of the Friedmann equation and then taken the square-root. This does not yield the same answer as taking $\overline{H}_\mathcal{D}$ using Eq.~(\ref{avH}) directly, which is the expectation value of the kinematical Hubble rate (but note that if we square Eq~(\ref{avH}), take the ensemble average, and then take the square-root we get the Hubble rate as  calculated directly from the Friedmann equation). The difference of course is the `ensemble-variance' of the Hubble rate, which may be defined by
\be
\text{Var}[H_\mathcal{D}]=\underbrace{\overline{H_\mathcal{D}^2}}_{\overline{[\text{Eq.~(\ref{Fried})}}]}-\underbrace{\overline{H}_\mathcal{D}^2}_{\overline{[\text{Eq.~(\ref{avH})}]}^2}.
\ee
When we write $\overline{[\text{Eq.~(\ref{avH})}]}^2$, this is developed to the correct perturbative order.

\begin{figure}[htbp]
\begin{center}
\includegraphics[width=0.5\textwidth]{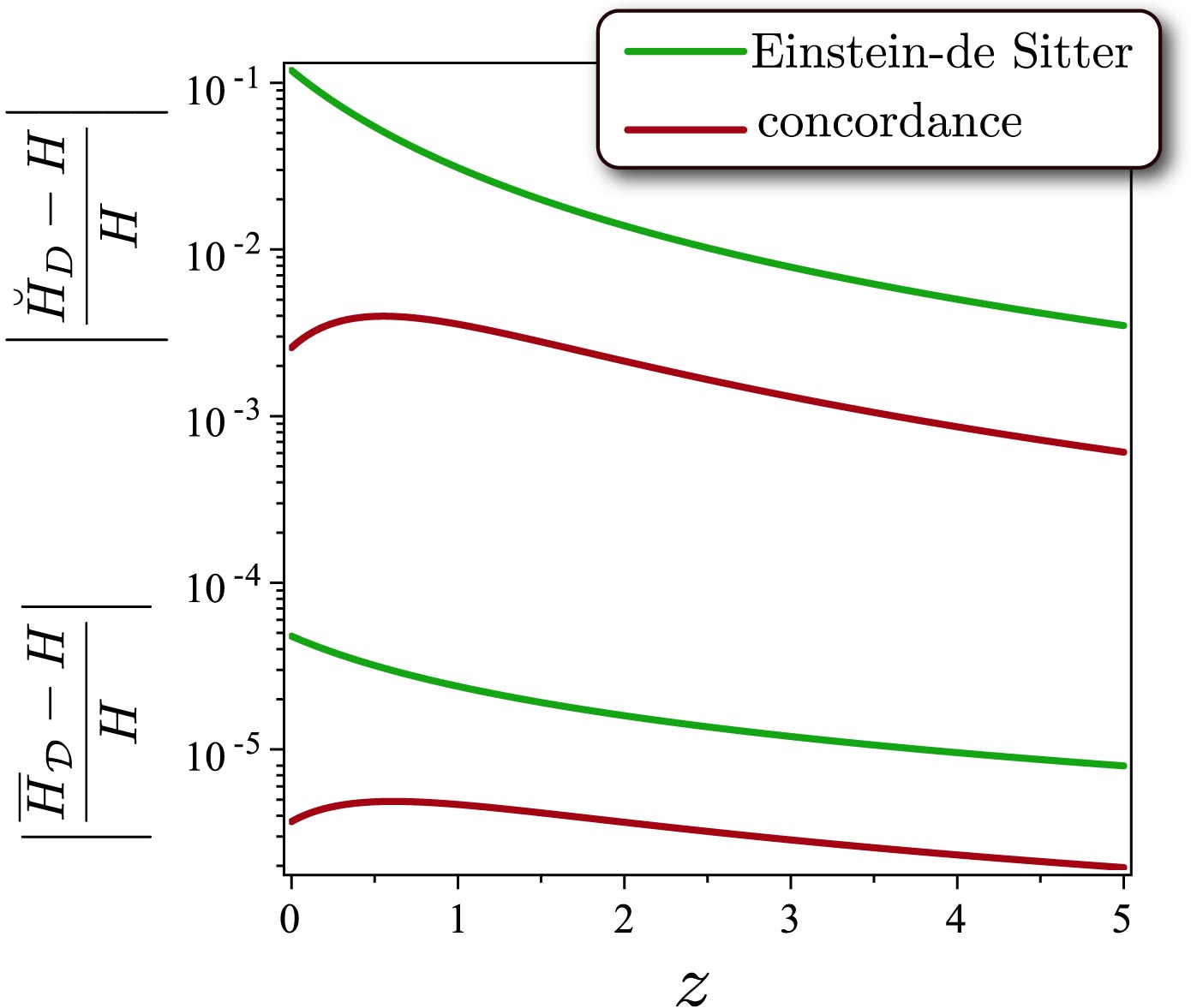}
\caption{The averaged Hubble rate as a function of redshift as a fractional change to the background Hubble rate. The top two curves show the Hubble rate $\breve H_\mathcal{D}$ calculated from the ensemble-averaged Friedmann equation, and the bottom two show the Hubble rate calculated directly, $\overline{H}_\mathcal{D}$.  Both concordance and EdS models are considered, and the domain size is set at the Silk scale, and the smoothing scale is set to zero.  }
\label{HR-z}
\end{center}
\end{figure}

\begin{figure}[htbp]
\begin{center}
\includegraphics[width=0.9\textwidth]{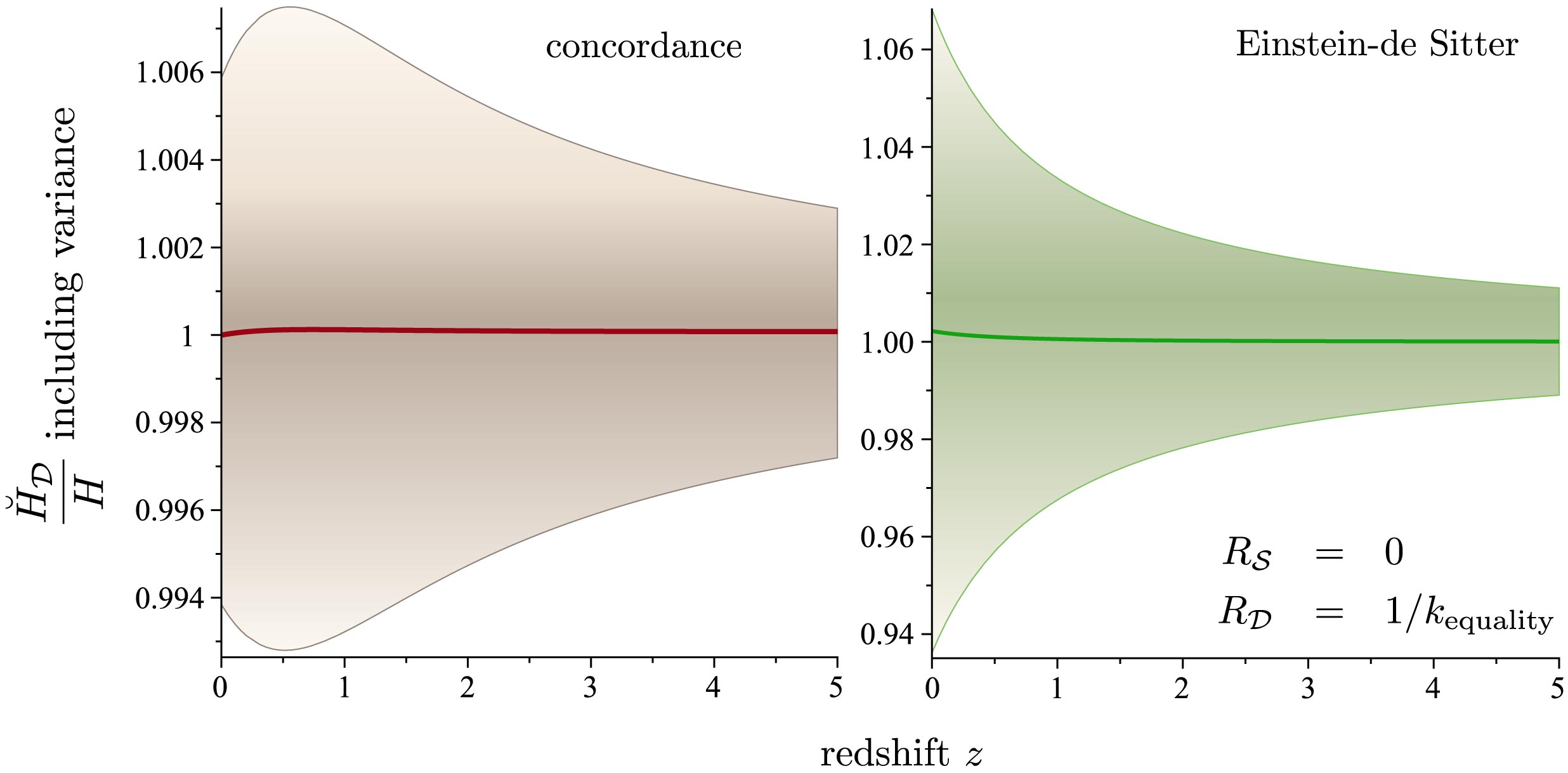}
\caption{Plots for $\breve H_\mathcal{D}$ with $R_\mathcal{D}=1/k_\text{equality}, R_\mathcal{S}=0$, with the variance included, as a function of redshift. }
\label{HR-z2}
\end{center}
\end{figure}

\begin{figure}[htbp]
\begin{center}
\includegraphics[width=0.6\textwidth]{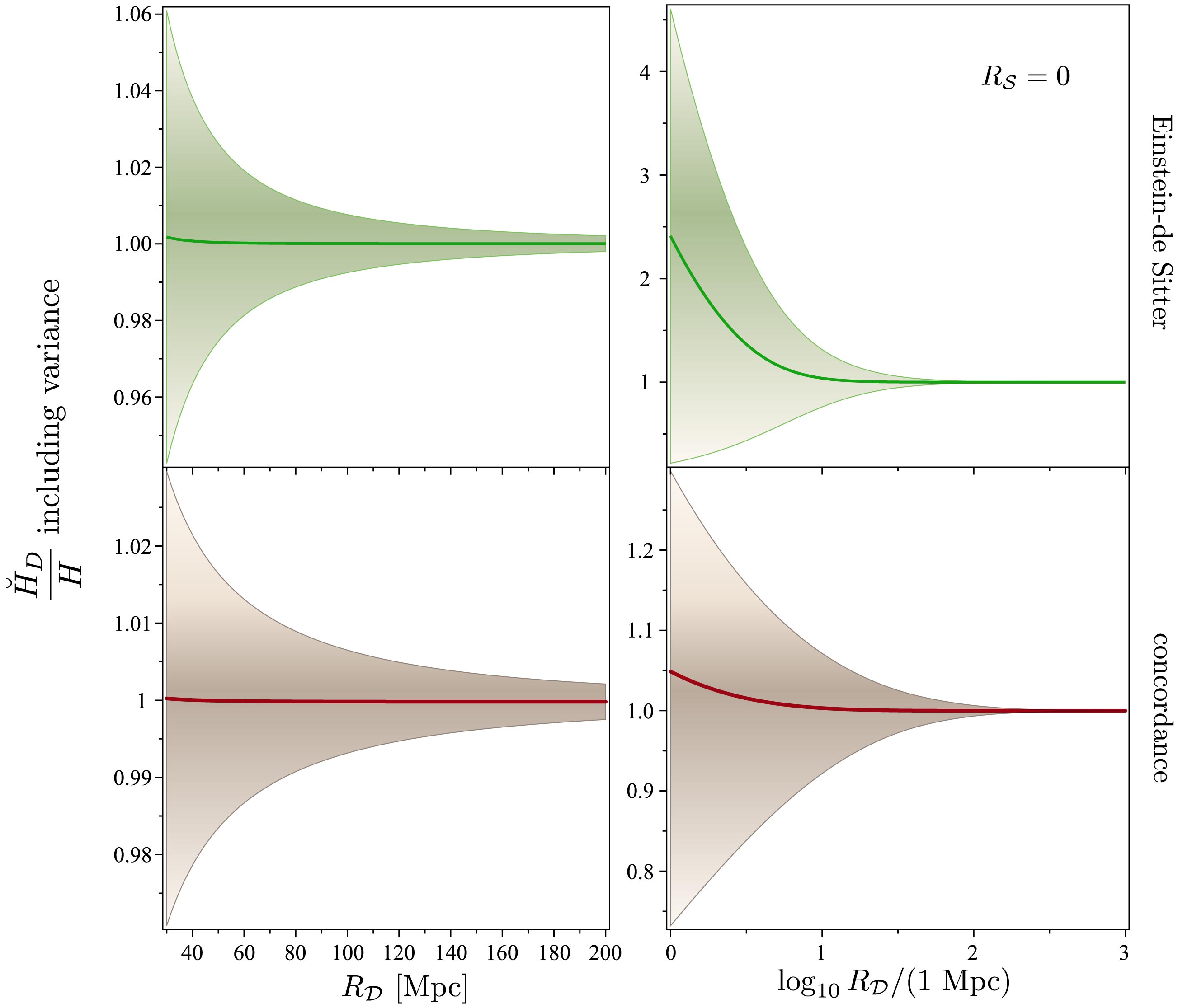}
\caption{The averaged Hubble rate today as a function of domain size, with the smoothing scale removed, $R_\mathcal{S}=0$, and with the variance included.}
\label{HD-RD}
\end{center}
\end{figure}

\begin{figure}[htbp]
\begin{center}
\includegraphics[width=0.4\textwidth]{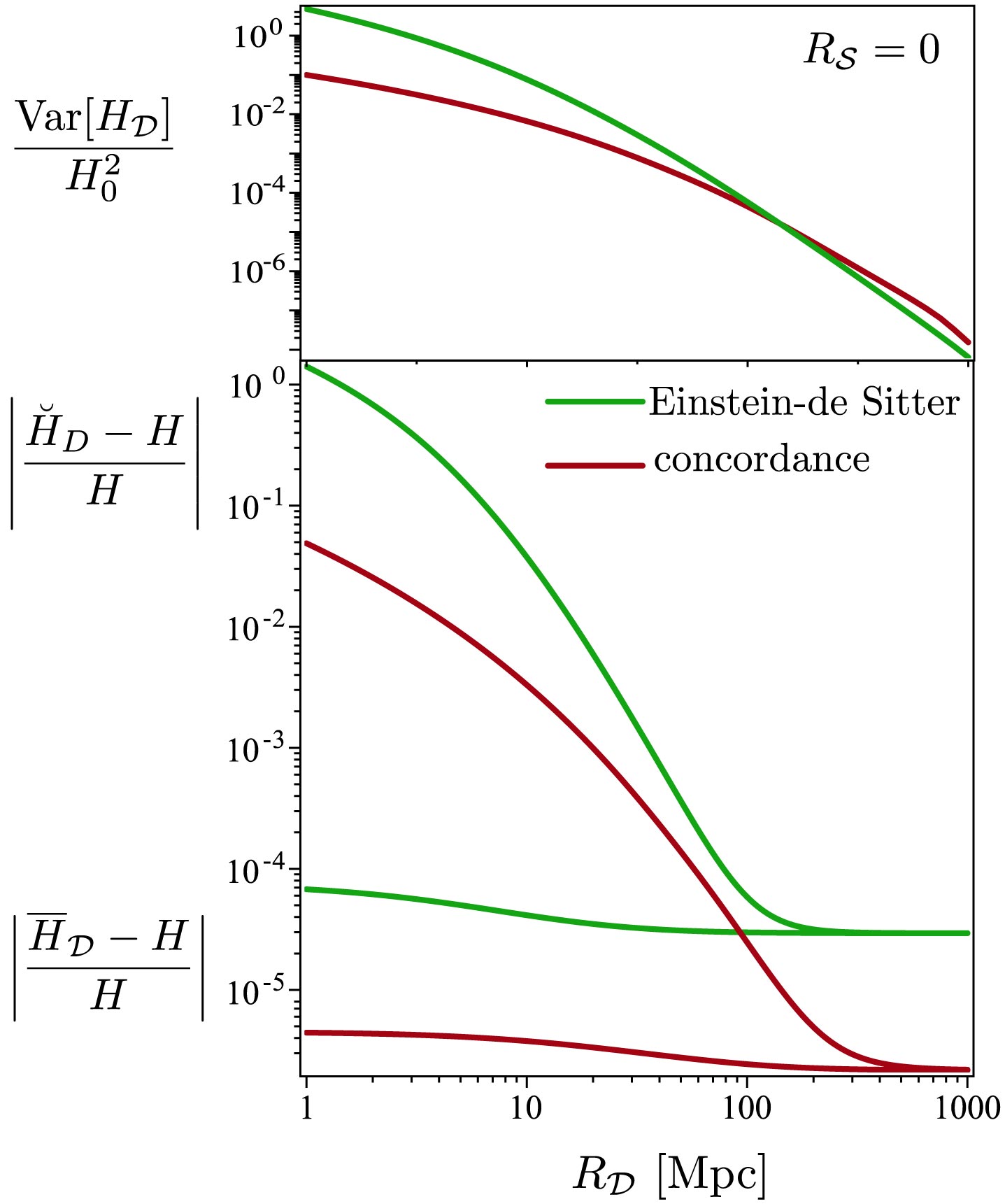}
\caption{
The averaged Hubble rates today as a function of domain size, and the variance (top), with the smoothing scale removed, $R_\mathcal{S}=0$. Note that both EdS and concordance models become independent of domain size when it becomes sufficiently large. Intriguingly this happens around $100-200$Mpc; at this scale, for both models, we also have $\breve H_\CD\sim\overline{H}_\CD$ and the variance reduces to the fiducial $10^{-5}$ level. }
\label{HR-today}
\end{center}
\end{figure}

\begin{figure}[htbp]
\begin{center}
\includegraphics[width=0.7\textwidth]{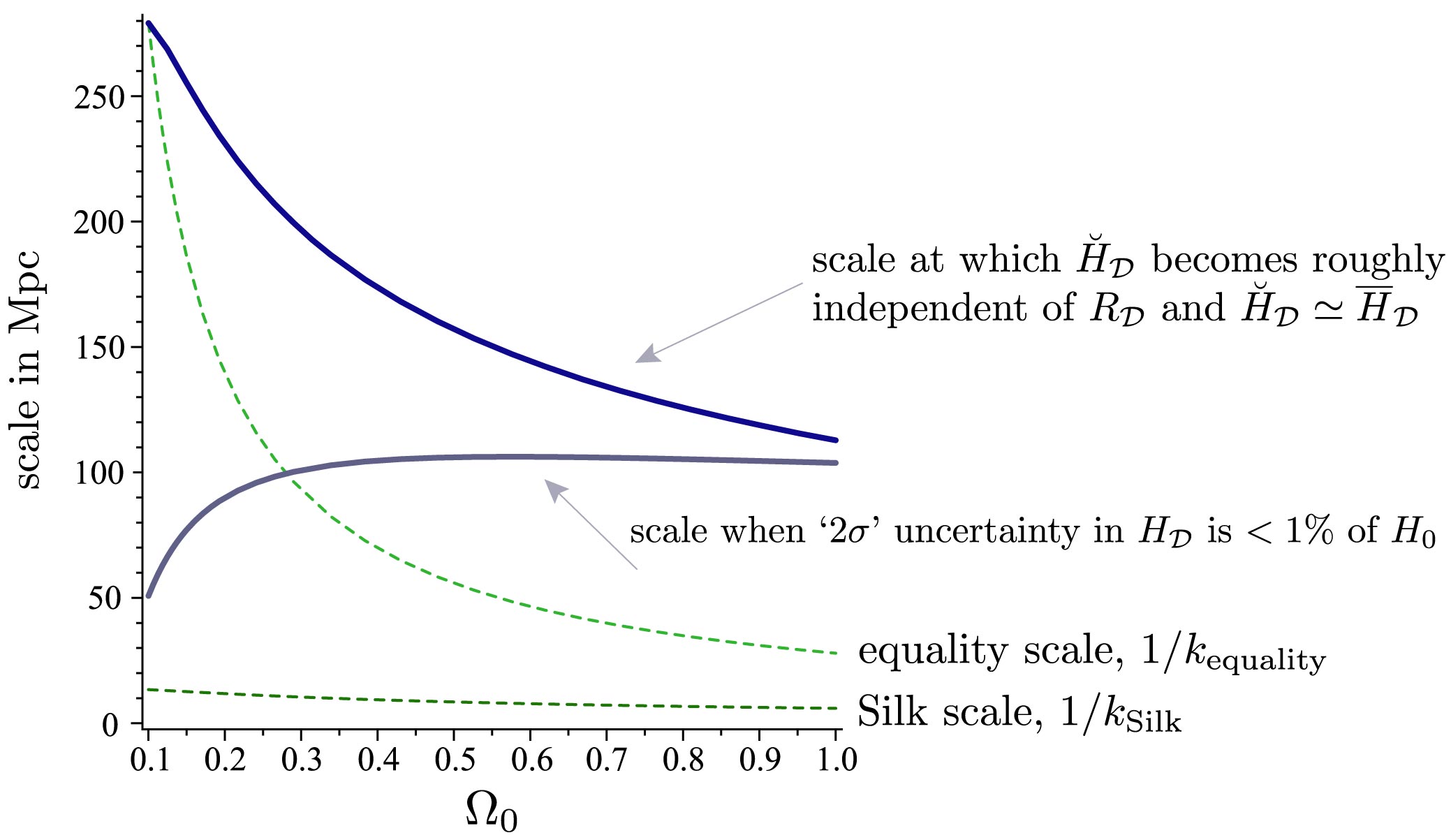}
\caption{Homogeneity scale in the averaged Hubble rate, as a function of present-day matter density. As the domain size is increased  the averaged Hubble rate settles down to a constant value which is different from $H_0$ (see Fig.~\ref{HR-today}). We may investigate this scale as a function of $\Omega_0$ by deciding when $H_\mathcal{D}$ is sufficiently independent io $R_\CD$ (here, $|\d\ln (\breve H_\CD/H_0-1)/\d\ln R_\CD|\lesssim\epsilon$~-- i.e., when the slope in the log-log plot of Fig.~\ref{HR-today} levels off). To illustrate, we have arbitrarily fixed this $\epsilon$ so that for  $\Omega_0=0.1$, the homogeneity scale coincides with the equality scale, but this curve can be moved up and down to taste.   Finally, another useful scale is when the variance becomes negligible: here we show the scale for which $2\sqrt{\text{Var}[H_\CD ]}=0.01H_0$. Note how the latter of these measures behaves in the opposite way to the scales intrinsic to the models~-- the equality and Silk scales. We have fixed the baryon fraction at 5\% here.}
\label{ScaledepRobust}
\end{center}
\end{figure}

In Figs.~\ref{HR-z}~-- \ref{ScaledepRobust} we show various aspects of the averaged Hubble rates. The backreaction effect grows during dust domination and decays after the dark energy transition. In Einstein-de Sitter, then, the backreaction effect is largest today, while in the concordance cosmology it peaks around $z\sim0.5$. The smoothing scale and most importantly the averaging domain size are crucial in deciding how large the effect of backreaction is.  We see also that whether we consider $\breve H_\mathcal{D}$ or $\overline{H}_\mathcal{D}$ matters considerably if the domain size is small; only if it is larger than a few hundred Mpc  do these agree.

As far as the Hubble rate is concerned there is no UV divergence in any of the integrals and so we have, for simplicity, set the smoothing scale to zero. In Figures~\ref{HD-RD} and~\ref{HR-today} we show the dependence on domain size on the Hubble rate today. It is largest for small  domain size, but we see that it levels off at a constant value for $R_\mathcal{D}$ bigger than a hundred Mpc, or thereabouts, and both Hubble rates are the same (i.e., the variance is sufficiently small). Recall that the domain scale only affects terms of the form $\overline{\<\cdots\>\<\cdots\>}$, so as $R_\mathcal{D}$ becomes large those terms disappear, leaving terms like $\overline{\<\cdots\>}$ which are independent of the domain scale. 

This can be interpreted by saying that there exists a scale which corresponds to the scale of homogeneity of the distribution of matter: there's an upper limit to domain size beyond which variables are fixed. This is explored further in Fig.~\ref{ScaledepRobust}, where we also show the scale beyond which uncertainty in $H_0$ drops to negligible levels.  The scale of homogeneity that comes out of the averaging process is similar to the equality scale, but behaves differently with $\Omega_0$, so the relation is not direct. 
In magnitude, this is similar to
 the scale of homogeneity inferred from large scale structure (LSS) surveys like SDSS \cite{SDSS,hogg}, but there are conflicting results in this area; see~\cite{SL1,SL2,SL3}, where it is found that homogeneity cannot be inferred from LSS surveys out to $100h^{-1}$Mpc. We are not able to say in this study whether back-reaction effects could account for such a discrepancy, but it is possible that the naive scale derived in linear theory is refined significantly when non-linear backreaction effects are properly accounted for. This deserves further investigation.

Thus we see that above a certain scale the backreaction leaves a small, scale-invariant, residue.  Above the scale of homogeneity, the backreaction effect becomes scale independent, but not at a zero value; in other words, the effective homogeneous model of the perturbed Universe that is obtained by averaging on scales bigger than a few hundred Mpc exhibits a `renormalized' background with non-zero backreaction. This is an important fact since we started with a given background, but we don't recover this background after averaging; that essentially means that {\it stricto sensu}, backreaction cannot be said to vanish on very large scales. In principle, this could provide a novel way of constraining the matter density by using measurements of this scale of homogeneity in large scale structure data.

 Our results are quantitatively comparable to those of \cite{lischwarz1} obtained in the synchronous gauge. In particular, the variance effect due to the backreaction exhibits the same scale behavior, with comparable amplitude. Our estimate of the variance is consistent with the variance in the Hubble rate calculated directly from the variance of peculiar velocities  in~\cite{wang}. This is important for future research aiming at 1\% uncertainty in measuring $H_0$~\cite{riess1,riess2}. We find comparable results for the change to the Hubble rate in  Refs.~\cite{Iain1,Iain2} where it was found that the backreaction in the Poisson gauge gives a change to the Hubble rate today $\sim10^{-5}$; we find about the same for the concordance model when averaging on Hubble scales, but we have shown that the backreaction can be large when averaging on sub-Hubble scales. Any difference can be attributed in part to the different definition used to calculate the efffective Hubble rate, and we also employ a different averaging scheme. We also include the second-order Bardeen potentials in our analysis, which were neglected in~\cite{Iain1,Iain2} though these are a sub-dominant contribution to the backreaction.

\subsection{The Raychaudhuri equation and deceleration parameter}

While we have seen the change due to backreaction in the Hubble rate are small, and, in particular, have no UV divergence, we can get further information about the effect of the backreaction from perturbations by looking at the Raychaudhuri equation. These are sourced in part by the new backreaction terms such as $\mathcal{K_D}$ and so on, but with many other quantities contributing. In terms of domain and smoothing scale, $\mathcal{K_D}$ diverges as $R_\mathcal{S}\to0$, and $\mathcal{Q_D-L_D}$ diverges as $R_\mathcal{D}\to0$; it becomes domain size-invariant beyond a few hundred Mpc. All the other backreaction terms are fairly independent of domain and smoothing scales.

We define an averaged deceleration parameter as
\be\label{q}
q_\mathcal{D}(z)=-\frac{1}{H_\mathcal{D}^2}\frac{\ddot{a}_\mathcal{D}}{a_\mathcal{D}},
\ee
where ${\ddot{a}_\mathcal{D}}/{a_\mathcal{D}}$ is given by the generalised Raychaudhuri equation, above. To calculate the ensemble average, we first  note that 
\be
\frac{\ddot{a}_\mathcal{D}}{a_\mathcal{D}}=H(z)^2\left(1-\frac{3}{2}\Omega_m(z)\right)+\delta^{(1)} R+\delta^{(2)}R
\ee
where $\delta^{(1)} R$ are all first-order terms such as $\<\Phi\>$ (which have zero ensemble average), $\delta^{(2)} R$ contains all second-order terms such as $\<\Phi^2\>$ and so on;
similarly for the Friedmann equation $H_\mathcal{D}^2=H^2+\delta^{(1)}F+\delta^{(2)}F$, where $\overline{\delta^{(1)}F}=0$. We then find
\be
\overline{q}_\mathcal{D}(z)=-1+\frac{3}{2}\Omega_m(z)-\frac{1}{H^2}\overline{\frac{\ddot{a}_\mathcal{D}}{a_\mathcal{D}}}+\left(1-\frac{3}{2}\Omega_m(z)\right)\frac{\overline{H_\mathcal{D}^2}}{H^2}+\frac{\overline{\delta^{(1)}R\,\delta^{(1)}F}}{H^4}. 
\ee
\begin{figure}[tbp]
\begin{center}
\includegraphics[width=0.98\textwidth]{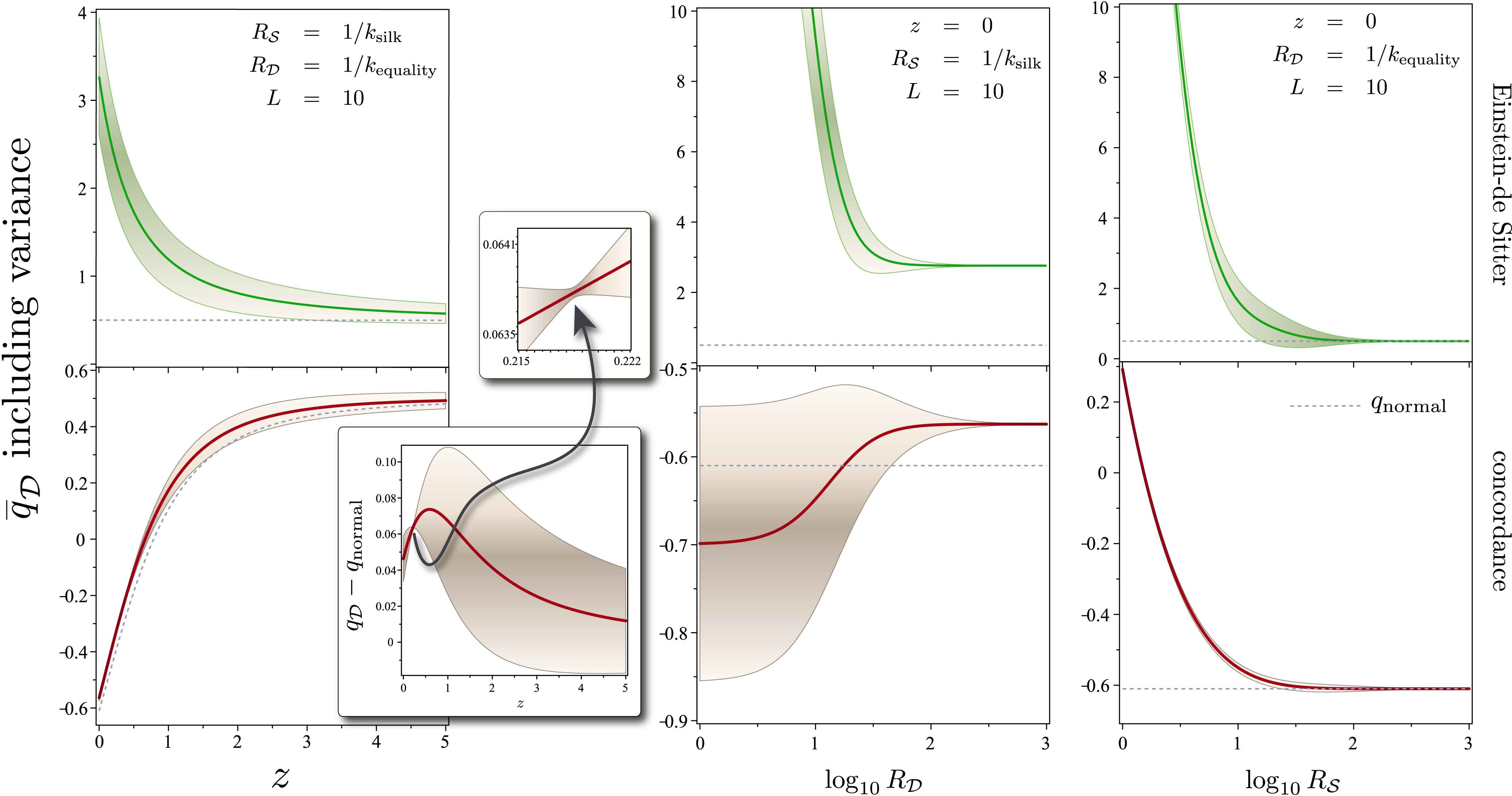}
\caption{The ensemble-averaged deceleration parameter and its variance, for the EdS and concordance models. We can see from the graphs on the left that the difference from normal behaviour grows during dust domination, and decays when the dark energy accelerated the expansion rate. Interestingly, the variance also starts to decrease then, reaching a minimum for $z\approx 0.22$. Note the divergence with the smoothing scale becomes irrelevant if it is bigger than tens of Mpc (right). The dependence on domain scale depends crucially on the model, however (middle), and no longer matters if $R_\mathcal{D}<R_\mathcal{S}$ (shown for the concordance model only, where the curves flatten off). }
\label{fig:q}
\end{center}
\end{figure}
\begin{figure}[t]
\begin{center}
\includegraphics[width=0.5\textwidth]{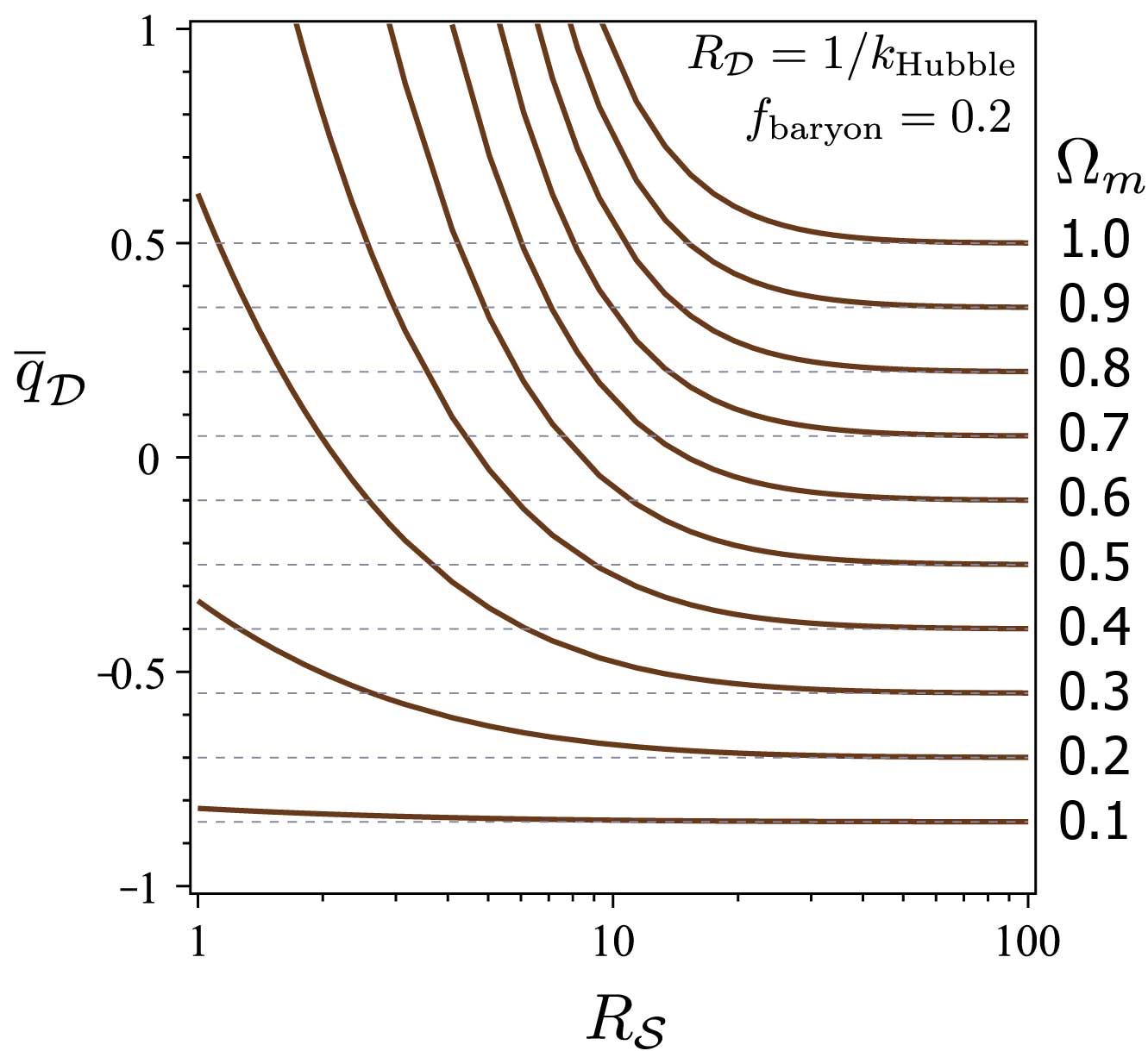}
\caption{The ensemble-averaged deceleration parameter today as a function of smoothing scale (in Mpc). This gives the residual contribution to the deceleration parameter for very large domains~-- clearly the smoothing scale is crucial, and suggests that higher-oder contributions might be necessary in order quantify the backreaction effect in $q_0$.}
\label{fig:q2}
\end{center}
\end{figure}
 (To high accuracy we get the same result if we just replace $H_\mathcal{D}^2$ with $\overline{H}_\mathcal{D}^2$ in Eq.~(\ref{q}).) To calculate the variance of $q_\mathcal{D}$ we must expand $q_\mathcal{D}^2$ to second-order to find $\overline{q_\mathcal{D}^2}$. We then find
\be
\text{Var}[q_\mathcal{D}]=H^{-4}\left\{\frac{1}{4}[2-3\Omega_m(z)]^2\overline{\delta^{(1)}F^2} - [2-3\Omega_m(z)] \overline{\delta^{(1)}F\delta^{(1)}R}+\overline{\delta^{(1)}R^2}\right\}.
\ee
In Figs.~\ref{fig:q} and~\ref{fig:q2} we explore the behaviour of the expectation value of the deceleration parameter, and its variance.

\section{Conclusion}

In this paper, we have calculated the effect of the backreaction arising from averaging a perturbed FLRW spacetime in the Poisson gauge consistenly up to second order in perturbation theory. Inclusion of a cosmological constant gives the analysis additional intricacy as we include the second-order Bardeen potentials, complications which have been ignored in previous studies. In particular they contain non-local terms which require some care to show that they vanish once an ensemble average is taken. We utilised an averaging procedure which has two elements: a spatial average over a typical domain, and an ensemble average over domains. The ensemble average gives the expectation value for a quantity averaged on a domain $\CD$, and the variance gives an intrinsic uncertainty to the expected value. We discussed in detail the changes that arise from averaging the Hubble parameter and the Friedmann equation, and the deceleration parameter. 

Our key results are:
\begin{itemize}
\item There exists a \emph{homogeneity scale} inherent in the averaging procedure which is not simply the equality scale for a general $\Omega_0$. As the domain size increases the averaged Hubble rate becomes independent of averaging scale, and the variance of the Hubble rate becomes negligible.  We also see a similar scale in the behaviour of the deceleration parameter. While this scale is comparable to the one inferred from large-scale structure surveys~\cite{SDSS,hogg}, we have seen that backreaction can give the homogeneity scale different to the equality scale somewhat. 
Further work could look into whether the homogeneity scale is calculated sufficiently accurately in linear theory, as backreaction may account for discrepancies in determining it from observations~\cite{SL1,SL2,SL3}.

\item For domains larger than the homogeneity scale, the backreaction does not go to zero; rather, there is a small residue left from the averaging, of order $10^{-5}$ for averaging the Hubble expansion on Hubble-scale patches. Thus, the background \emph{is} renormalised by structure formation, and so the `background' at late times is not the same as the background at the end of inflation.

\item It is not clear exactly how large the change to the deceleration parameter is.  A seemingly sensible and conservative choice of smoothing scale~-- the Silk scale~-- gives a sizeable change to the deceleration parameter, of the order of 10\% (for our concordance model~-- more as $\Omega_0$ increases), even when averaging above the homogeneity scale. If it is this large, this effect could be critical in determining the nature of dark energy. Furthermore, the deceleration parameter is pushed to more positive values by backreaction. The existence of a UV divergence in the effective Raychaudhuri equation implies that to quantify this properly, higher-order perturbation theory might be required. However, it has been argued~\cite{notari} that going to higher-order in perturbation theory may make the UV divergence worse, not better.\footnote{The problem is that at higher orders in perturbation theory higher derivatives of the first-order Bardeen potential appear, which, on the face of it, may make any UV divergences worse; it may even be argued that perturbation theory is not convergent on the basis of this~\cite{notari,kolb-pc}. Our models of the universe may not be analytic. Alternatively, it may be more like convergence in the power series of $e^x$ as we increase $x$: we need more terms for convergence, each bigger than the last. In this case, third- or fourth- order perturbation theory may help extend the region of validity of these calculations, but perhaps only a little.}

\item Backreaction also introduces an uncertainty in the value of the Hubble rate and deceleration parameter which manifests itself through a variance in the value in different averaged regions. This variance has been estimated here and is found to be quite large if the averaging scale is below the homogeneity scale discussed above (around a few percent for small averaging scales), in approximate agreement with estimations previously made in the synchronous gauge \cite{lischwarz1}, as well as estimates made independently of the relativistic averaging scheme~\cite{wang}. This may be an important consideration for observationally determining cosmological parameters such as $H_0$ to accuracy higher than a few percent.

\end{itemize}

There exists an IR divergence inherent in this analysis. This means that all the modes bigger than the one corresponding to the averaging scale contribute to the backreaction effect, irrespective to their size, for a scale-invariant spectrum. In particular, the Hubble radius does not appear as a natural cut-off, and the smallness of the effect noted above only comes from the introduction of an arbitrary cut-off: by pushing this cut-off towards higher values, the backreaction effect is larger, though only logarithmically with scale. For a red spectrum, however, this divergence is power-law so could be important. In contrast to~\cite{KMNR} we find that this divergence appears in the backreaction effect directly and not just the variance. The longest wavelength mode should perhaps be set by the start of inflation, though it is by no means clear exactly how to implement this properly. 

The assumption of a primordial  spectrum of adiabatic Gaussian perturbations was also critical in this analysis, and all conclusions we present are reliant on this fact. In particular, because the first-order contribution to the backreaction vanishes for Gaussian perturbations, non-Gaussian initial perturbations may produce a much larger backreaction effect, of the order of the square-root of the variance we have presented here. This is an important outstanding issue to explore.

\acknowledgements

We would like to thank Iain Brown, Pier-Stefano Corasaniti, Christiano Germani, Alex Hamilton, Rocky Kolb, Kazuya Koyama, Roy Maartens, Alessio Notari, Francesco Sylos Labini and Yun Wang for discussions and/or comments, and especially Ruth Durrer for the same and for considerable help with all the ensemble averaging methods. CC is supported by the NRF (South Africa); KA and JL are supported by the Claude Leon Foundation (South Africa).

\end{document}